\documentclass[12pt,onecolumn]{IEEEtran}
\usepackage{tikz}
\usepackage{scalerel}
\usepackage{times}
\usepackage{relsize}

\usepackage{enumitem}
\usepackage{float}
\usepackage{mathtools}
\usepackage{graphicx}
\usepackage{subfig}
\usepackage{verbatim}
\usepackage{graphicx,float,latexsym,amssymb,amsfonts,amsmath,amstext,times,epsfig,setspace}
\usepackage{algorithm}
\usepackage[utf8]{inputenc}
\usepackage[noend]{algpseudocode}
\usepackage{fancyhdr}
\usepackage{comment}
\usepackage{stfloats}
\usepackage{tabu}
\usepackage{booktabs}
\usepackage{amsmath}
\usepackage{dsfont}
\usepackage{comment}
\usepackage[noend]{algpseudocode}
\renewcommand{\algorithmiccomment}[1]{\bgroup\hfill\scriptsize//~#1\egroup}

\usepackage{pgfplots}
\newcommand{\btikzpicture}{\begin{tikzpicture}}
\newcommand{\etikzpicture}{\end{tikzpicture}}
\pgfmathsetmacro{\nudge}{0.04}
\pgfmathsetmacro{\gh}{0.05}
\pgfmathsetmacro{\ANb}{0.65}
\pgfmathsetmacro{\VNb}{-0.6}
\pgfplotsset{compat=1.16}

\DeclareMathOperator*{\argmin}{arg\,min}

\newcommand{\srcdata}{source data}

\newcommand{\Ndata}{N}

\newcommand{\erate}{{\cal E}}
\newcommand{\rrate}{{\cal R}}

\newcommand{\clen}{{\rm clen}}

\newcommand{\bitval}{bit}

\newtheorem{theorem}{Theorem}[section]
\newtheorem{lemma}[theorem]{Lemma}

\newtheorem{proposition}[theorem]{Proposition}
\newtheorem{corollary}[theorem]{Corollary}

\newcommand{\qed}{\nobreak \ifvmode \relax \else
      \ifdim\lastskip<1.5em \hskip-\lastskip
      \hskip1.5em plus0em minus0.5em \fi \nobreak
      \vrule height0.75em width0.5em depth0.25em\fi}

\newcommand{\reals}{\ensuremath{\mathbb{R}}}

\newcommand{\expectation}{\ensuremath{\mathbb{E}}}

\newcommand{\prob}{\ensuremath{\mathbb{P}}}

\newcommand{\nobject}{\ensuremath{n_{\text obj}}}

\newcommand{\Efragments}[2]{\ensuremath{{\cal F}(#1,#2)}}

\newcommand{\node}{\ensuremath{{\mathfrak  n}}}

\newcommand{\Tsysrep}{\ensuremath{T_{\text{tot}}}}

\newcommand{\mtrans}{v}
\newcommand{\failtime}{\tau_F}

\newcommand{\survive}{{\cal S}}
\newcommand{\launchtime}{\tau_L}
\newcommand{\launchid}{I_L}

\newcommand{\settletime}{\tau_s}

\newcommand{\gstar}{\boldsymbol{\delta}}
\newcommand{\bgstar}{\overline{\gstar}}
\newcommand{\betastar}{\beta_{\gstar}}
\newcommand{\tfail}{s}

\newcommand{\Tnode}{\Delta_t}%{\delta\! T_\node}

\newcommand{\nvfobj}{\kappa}

\newcommand{\kcode}{k_c}
\newcommand{\GAP}{2\gstar}

\begin{document}
%\pagenumbering{gobble}

\title{Distributed storage algorithms \\ with optimal tradeoffs}
\thispagestyle{fancy}
%\lhead{\small IEEE TRANSACTIONS ON INFORMATION THEORY,~Vol.~?, No.~?, May~2015}
\author{
\IEEEauthorblockN{
Michael~Luby\IEEEauthorrefmark{1}\IEEEmembership{~IEEE~Fellow,~ACM~Fellow},
Thomas~Richardson\IEEEauthorrefmark{2}\IEEEmembership{~IEEE~Fellow}
}\\
\IEEEauthorblockA{
\IEEEauthorrefmark{1}%
BitRipple, Inc
}
\IEEEauthorblockA{
\IEEEauthorrefmark{2}%
Qualcomm Technologies Inc.
}
\thanks{Portions of this work were done while the first author was with Qualcomm Technologies, Inc. Portions of this work was supported by the National Science Foundation under grant 1936572. BitRipple, Inc (luby@bitripple.com) and International Computer Science Institute (luby@icsi.berkeley.edu).}
}
%\thanks{Revised draft: \today}
%}

%\fancyhead[RO,LE]{\thepage}
%\fancyhead[RE]{\small IEEE TRANSACTIONS ON INFORMATION THEORY, \date{The date}{\today}}
%\fancyhead[LO]{\small Reduced draft}
%\fancyfoot[L]{\em Reduced draft: \today}
\renewcommand{\headrulewidth}{0pt}% disable the underline of the header part

\maketitle

\begin{abstract}

One of the primary objectives of a distributed storage system is to reliably store large
amounts of \srcdata\ for long durations using a large number $\Ndata$
of unreliable storage nodes, each with $\clen$ \bitval s of storage capacity.  
Storage nodes fail randomly over time and are replaced with nodes of equal capacity initialized to zeroes, and thus \bitval s are erased at some rate $\erate$. 
To maintain recoverability of the \srcdata,
a repairer continually reads data over a network from nodes at a rate $\rrate$, 
and generates and writes data to nodes based on the read data.

The {\em distributed storage \srcdata\ capacity} is the maximum
amount of \srcdata\ that can be reliably stored for long periods of time.
The research described in~\cite{Luby19} shows that asymptotically the 
distributed storage \srcdata\ capacity is at most
\begin{equation}
\label{fundamental eq}
\left(1-\frac{\erate}{2 \cdot \rrate}\right) \cdot \Ndata \cdot \clen
\end{equation}
as $\Ndata$ and $\rrate$ grow.

In this work we introduce and analyze algorithms such that asymptotically the 
distributed storage \srcdata\ capacity is at least Equation~\eqref{fundamental eq}.
Thus, Equation~\eqref{fundamental eq} expresses a fundamental trade-off
between network traffic and storage overhead to reliably store \srcdata.
\end{abstract}


\IEEEpeerreviewmaketitle

\section{The Basic Liquid System}

In this section we review the basic liquid storage system model, as developed in \cite{Luby16}, and indicate some of the extensions introduced in this paper.
We establish the essential mathematical framework of liquid storage, and prepare for variations that improve on certain characteristics, in particular read bandwidth requirements for repair.
Many practical details discussed in \cite{Luby16} are outside the scope of this paper and, for purposes of exposition, we make certain simplifying assumptions in the model.

In the storage system models we consider a total of \nobject~ equal-sized data objects are stored.
Each object's source data is partitioned in $\kcode$ equal size data fragments and
an MDS\footnote{In \cite{Luby16} nearly MDS codes, RaptorQ codes, were proposed for their scalability and complexity properties.  Here we assume MDS codes largely for convenience, in a practical implementation of the proposed schemes RaptorQ would still likely be the best practical choice.} erasure correcting code is used to generate as many as $\kcode+r$ fragments corresponding to a length $\kcode+r$ code. The code may or may not be systematic.
In the basic liquid system the code length, $\kcode+r,$ equals $N,$ the number of storage nodes used in the system
and each storage node is uniquely associated to a symbol in the error correcting code.  Data fragments corresponding to a particular symbol are stored on the associated node.  
In the basic liquid system this association is exclusive, the node stores all fragments associated to the code symbol and no others.
In the extensions considered in this paper we may have $\kcode+r > N$ and the total number of fragments generated upon object repair can vary and may be less than 
$\kcode+r$.  Nodes will generally be associated to code symbols in the code but fragments associated to other code symbols may also be stored on a node.

We define $\beta= \frac{N-\kcode}{N}$ as the {\em overhead} of the storage system.  
The stored source data objects comprise $(1-\beta)$ of the storage system capacity and 
the remaining fraction $\beta$ of the storage capacity is used to provide resiliency against data loss due to node failure.

We say that a fragment is {\em intact} at time $t$ if it is stored on a node and can be read at that time.
%A fragment is associated to an object and to a symbol of the code.
When a storage node fails all fragments stored on it are lost.  
The node is then replaced with a new empty node, and we assume that this happens instantaneously.\footnote{Part of the appeal of liquid storage is that repair, including node replacement, can be delayed without signiificantly affecting the risk of data loss.  We assume instantaneous node replacement largely as a mathematical convenience.}
Thus node failure is functionally equivalent to node erasure and we will often refer to fragment loss as fragment erasure.
In the basic liquid system the replacement node is assigned the same code symbol as the node it replaces.  
In the systems developed in this paper this is no longer the case.
Instead we will assume an ordered list of symbols and when a node fails its associated symbol is placed at the bottom of the list and
the new node is associated to the first unassociated symbol on the list.
%The repair efficiency of the liquid storage systems stems in large part from delaying the regeneration of lost fragments until such time that the regeneration can be performed efficiently.

In the basic liquid system, objects are repaired serially in a fixed cyclic order.  When an object is repaired, $\kcode$ fragments are read\footnote{Practical systems would likely read, or at least access, slightly more than $k$ fragments to reduce latency due to straggler nodes.  In the case of non-MDS codes more than $\kcode$ fragments might be needed.  In this paper we ignore these marginal effects.}, which are sufficient to reconstruct the data, and any erased fragment is (re)generated and written to its associated node, i.e., the node associated to the corresponding code symbol.  
Hence, immediately upon repair, all $N$ of an object's fragments are intact. 
We treat object repair as an atomic event, i.e., we do not concern ourselves with partial repairs of objects being interrupted by node failure.
We associate the time of this atomic event with the completion of the repair.
{\em Repair efficiency} is related to the number of erased fragments regenerated per object repair, and regenerating a large number of erased fragments makes more efficient use of $\kcode$-fragment read than regenerating a smaller number. 
In the basic liquid system, the maximum number of fragments that can be regenerated is $N-\kcode$ since if more than $N-\kcode$ fragments are erased then the data object is not recoverable.
For efficient repair it is desirable to operate the system so that then number of fragments regenerated is near this maximum.
This aim must be balanced against the risk of loosing data, and some margin must be maintained.  
Exploiting the law of large numbers, larger systems can tolerate smaller relative margins.
In the systems considered in this paper objects will not necessarily be repaired in strictly cyclic order, but cyclic repair will be loosely followed.
We will also generally treat repair as an atomic event, but this atomic event may involve more than one object repair.
In a typical object repair we will regenerate more than $N-\kcode$ fragments, thereby increasing repair efficiency.
Some of those fragments will correspond to code symbols not yet associated to an actual storage node in the system (they may instead be associated to virtual nodes) and those
fragments will be temporarily stored in other locations until the associated node is physically introduced into the system, at which point those
fragments are moved (copied) to that node.

In this paper we often assume a fixed rate of object repair since we are interested in asymptotic performance limits that use minimal repair resources.
This rate of repair is characterized by $ \Tsysrep$ which denotes the time required to repair all objects once each.
In the basic liquid system this is equivalent to  the time between repairs for a fixed object since the repair cycles through the objects but this will not hold 
in the extended systems.
We generally assume that the time to failure of a node is an exponentially distributed random variable of known rate.
With fixed system size, this assumption is equivalent to a Poisson node failure process.
In practice the node failure rate may not be precisely known and the Poisson assumption may not be valid.
In \cite{Luby16} a feedback regulator was described that modulates the repair rate as a function of the observed node failure process.
To maintain repair efficiency the regulator attempts to steer the number of repaired fragments toward some target, the choice of which balances efficiency against the risk of data loss.
We will indicate how the technique can be extended to the models considered here.
%a feedback regulator was described that modulates the repair rate as a function of the node failure realization.
%Such a regulator can be adapted to the schemes propsed here, and we will give some indication of the required changes to the system.

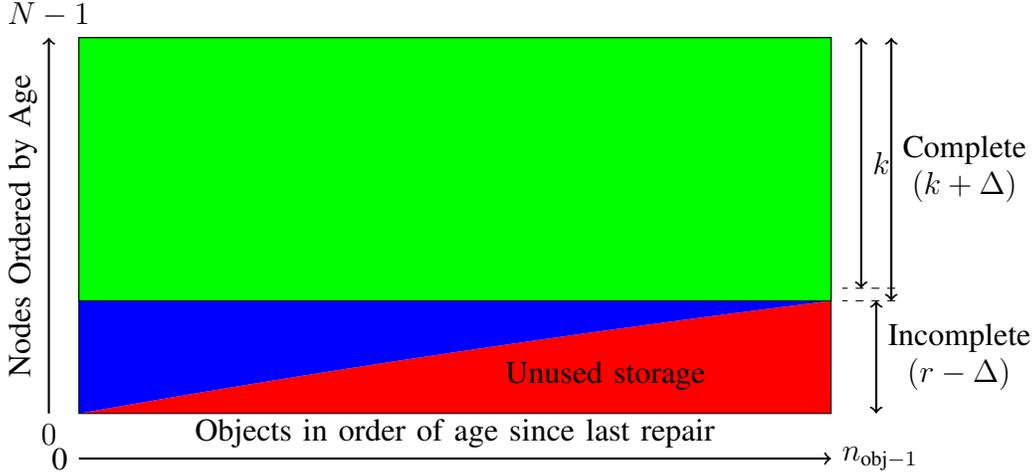
\begin{figure}
\begin{center}
\begin{tikzpicture}[x=10cm,y=5cm, declare function={ ff(\x) =  1.0-exp(-0.3566*\x); } ]
%  \fill[red, domain=0:10, variable=\x]   (0,0) -- plot ({\x},  {0.09*\x})  --(10,0) --cycle;
% \fill[blue, domain=0:10, variable=\x]   (0,0.9) -- plot ({\x},  {0.09*\x})  --cycle;
  \fill[red, domain=0:1, variable=\x]   (0,0) -- plot ({\x},  { ff(\x)})  --(1,0) --cycle;
  \node at (0.7,0.1) {Unused storage};
 \fill[blue, domain=0:1, variable=\x]  (0,0.3) -- plot ({\x},  { ff(\x)})  --cycle;
 
\draw (0,0) rectangle (1,1);
\draw[fill=green] (0,0.3) rectangle (1,1);

\node[left] (A) at (0,-3*\nudge) { 0 };
  \node[right] (B) at (1,-3*\nudge) { $n_{\text{obj}-1}$ };
  \draw[thick,->] (A)--(B) node[midway,above] { Objects in order of age since last repair };
  
\node[above]  (TR) at (1+2*\nudge,1) { };
  \node[below] (REP) at (1+2*\nudge,0.3) { };
   \node[above]  (TRl) at (1+\nudge,1) { };
  \node[below] (REPl) at (1+\nudge,0.333) { };
  \draw[thick,<->] (TR)--(REP) node[midway,right,align=center] { Complete \\ $(k+\Delta)$ };
   \draw[thick,<->] (TRl)--(REPl) node[midway,right] { $k$ };
\node (v2) at (1,0.3) {};
\node (v1) at (1.1,0.3) {};
\draw[dashed]  (v1) edge (v2);
\node (v2l) at (1,0.3333) {};
\node (v1l) at (1.1,0.3333) {};
\draw[dashed]  (v1l) edge (v2l);
\node[below] (v3) at (-\nudge,0) { $0$};
\node[above] (v4) at (-\nudge,1) { $N-1$};
\draw[thick,->]  (v3) -- (v4) node[midway,above,rotate=90] { Nodes Ordered by Age };
\node[above] (v6) at (1+1.5*\nudge,0.3) {};
\node[below] (v5) at (1+1.5*\nudge,0) {};
\draw[thick,<->]  (v5) -- (v6) node[midway,right,align=center] { Incomplete \\ $(r-\Delta)$ };
\end{tikzpicture}
\end{center}
\caption
{\label{figbasicrepairqueue}Visualization of basic liquid repair queue.  The head of the queue is on the right and the tail on the left.
Objects at the head of the queue are repaired and then move to the tail.  Hence, the cyclic order of the objects is invariant.}
\end{figure}

It is convenient to introduce a two dimensional visualization of the repair process in which one axis (the $x$ axis) represents ordered objects in the system and the other axis (the $y$ axis) represents the nodes.  In this visualization storage capacity is faithfully represented as area, see Fig. \ref{figbasicrepairqueue}.
Since objects are effectively queued for repair, we will also refer to this as the repair queue.
At any time the {\em position} of an object in the repair queue is the number of objects {\em behind} it in the queue.
In the basic liquid system, with its strict cyclic repair order, this is identical with the number of objects that have been repaired since the given object was last repaired.
Under this interpretation, an object's position is an integer in $0,...,\nobject-1.$
It is sometimes convenient to introduce a real variable $x \in [0,1)$ to represent position in a scale invariant manner.
By {\em the object at position $x$} we will mean the object at integer position $\lfloor x \nobject \rfloor$
and we will use both "object-position" and "$x$-position" to refer to position in the repair queue.
To avoid distracting complications we will often tacitly assume operation in the liquid limit, by which we mean the limit of an infinite number of objects.
In particular, in that limit any distinction between $x \nobject$ and  $\lfloor x \nobject \rfloor$ disappears.
%Alternatively, the fractional portion of a position may sometimes be used to represent the partial time elapsed since the last object repair.
We represent the storage nodes as equally spaced on the $y$ axis, where each node occupies an interval of height $1/N$ so that the storage capacity of the
node corresponds with a rectangular strip of height $1/N$ along the $y$-axis, and length $1$ along the $x$-axis.
Thus, the vertical axis is normalized to have $y \in [0,1).$   
In this visualization, the nodes are generally ordered by their age. % which is the time since their last erasure.  
If the node associated to the vertical segments $[k/N,{(k+1)}/N],$ fails, then the $y$ axis location of
nodes $0,...,k-1$ are each increased by $1/N$ and the empty replacement node is placed in the strip in $y$-position $[0,1/N).$
The repair process then ensures that the set of objects which store intact fragments on a node is non-decreasing (by inclusion) in $y.$
Whereas in the basic liquid system a replacement node immediately starts storing regenerated fragments, in the systems considered in this paper use of the replacement node for storage may be delayed until a certain transition in the repair process occurs.  Until that time the replacement node does not actively function in the system and storage on that node will be virtual.

Assume an object is repaired at time $t=0$ and the object is then placed at the tail of the repair queue.
After some additional time $t$ has elapsed some of the object's fragments may be lost due to node failure.
Let $\Efragments{x}{t}$ denote the set of erased fragment symbols for the object in position $x.$ 
%at time $t$
%and let $\Ifragments{x}{t}$ denote the set of intact fragments.
If $x_1 \le x_2$ then we have $\Efragments{x_1}{t} \subset \Efragments{x_2}{t}.$
This nested fragment ordering is central to operation in the basic liquid system.
The systems in this paper will largely preserve this property, but some deviation will occur.

Let us define the function $f(x,t) = \frac{1}{N}|\Efragments{x}{t}|.$
Since there are initially $N$ fragments, one each on a node, and node lifetimes are independent and exponentially distributed with rate $\lambda,$
the distribution of the number of erased fragments at time $t$ is binomial.  In particular, the probability that $k$ fragments have been erased by time $t$ is
$\binom{N}{k} (1-e^{-\lambda t})^k (e^{-\lambda t})^{N-k}.$
Assuming  $t< \Tsysrep,$ the object will be in $x$-position $t/\Tsysrep,$ and we see that the expected number of erased fragments for an object in position $x$
is $N \expectation f(x,t) = N (1-e^{-\lambda \Tsysrep x}).$  

%Assuming a fixed repair rate and Poisson node failures in the limit of large $N$ with fixed code rate the function $f$ concentrates around its expected value which takes the form
%\(
%f(x,t) = 1 - e^{-\lambda \Tsysrep x}
%\)
%where $\Tsysrep$ is the time required to repair all objects once each and $\lambda$ is the node failure rate \cite{Luby16}.

The repair efficiency of the basic liquid system stems from the number of fragments regenerated upon repair.  
For an object repaired at time $t$ the number of repair fragments regenerated is $f(1,t) N.$
Under our current assumptions the probability that this takes the value $\ell$ is 
$q(\ell) = \binom{N}{\ell} (1-e^{-\lambda \Tsysrep})^k (e^{-\lambda \Tsysrep})^{N-\ell}$
and has an expected value of $N (1-e^{-\lambda \Tsysrep}).$  
Data loss occurs if $\ell>r.$  In particular $\Tsysrep$ should be chosen so that $N (1-e^{-\lambda \Tsysrep}) < r.$
In \cite{Luby16} the bound
\[
\text{MTTDL} \ge \frac{1}{\lambda N (1-\beta) q(r)} - \frac{\Tsysrep}{\sum_{j=0}^r q(j)}
\]
was shown, where $\text{MTTDL}$ denotes the mean time to data loss assuming a perfect (all fragments intact) initial state.
For an appropriate choice of $\Tsysrep$ we have $\sum_{j=0}^r q(j) \simeq 1$ and $q(r) \ll 1.$
The $\text{MTTDL}$ from a typical state is essentially the same as that from a perfect state \cite{Luby16}.

%If the repair rate is sustained at too high a rate then $f(x,t)$ drops to a low value and the repair efficiency drops.
%Intuitively, one might consider suspending repair or lowering the repair rate to prevent this drop.  
%Lowering the rate over an extended period, however, runs the risk of data loss and rapid response in the repair rate has the potential to produce instabilities.
\subsection{Asymptotic Repair Read Rate}
In the limit $N\rightarrow \infty$ we can choose $\Tsysrep(N)$ such that $(1-e^{-\lambda \Tsysrep(N)})\rightarrow \beta$ and
$\text{MTTDL} \rightarrow\infty.$   In this limit the system achieves its maximum per-repair  efficiency with an object repair rate given by
 $\Tsysrep = (1-e^{-\lambda \Tsysrep(N)})\rightarrow -\lambda^{-1} \ln (1-\beta).$
In \cite{Luby19} a lower bound on data read rates was established that for small $\beta$ can be written as $\lambda \Tsysrep \ge \frac{\beta}{2}.$ 
Compared to this bound the read rate in the basic liquid system is larger, essentially by a factor of $2.$

An evident cause of this gap is the typically unused portion of the overhead storage capacity in the basic liquid system, see Fig. \ref{figbasicrepairqueue}.
The area under the curve $y=f(x,t)$ represents unused storage capacity and in the asymptotic limit discussed above this unused capacity
exceeds $\frac{\beta}{2}.$  If in the basic liquid system we excluded the
unused portion of the overhead storage capacity in the accounting of the overhead, then the lower read rate bound would be achieved
to within second order in $\beta.$
This observation is the key to the storage schemes presented in this paper where we aim to achieve full storage utilization.

\section{ Partially Virtualized Repair Queue Schemes}

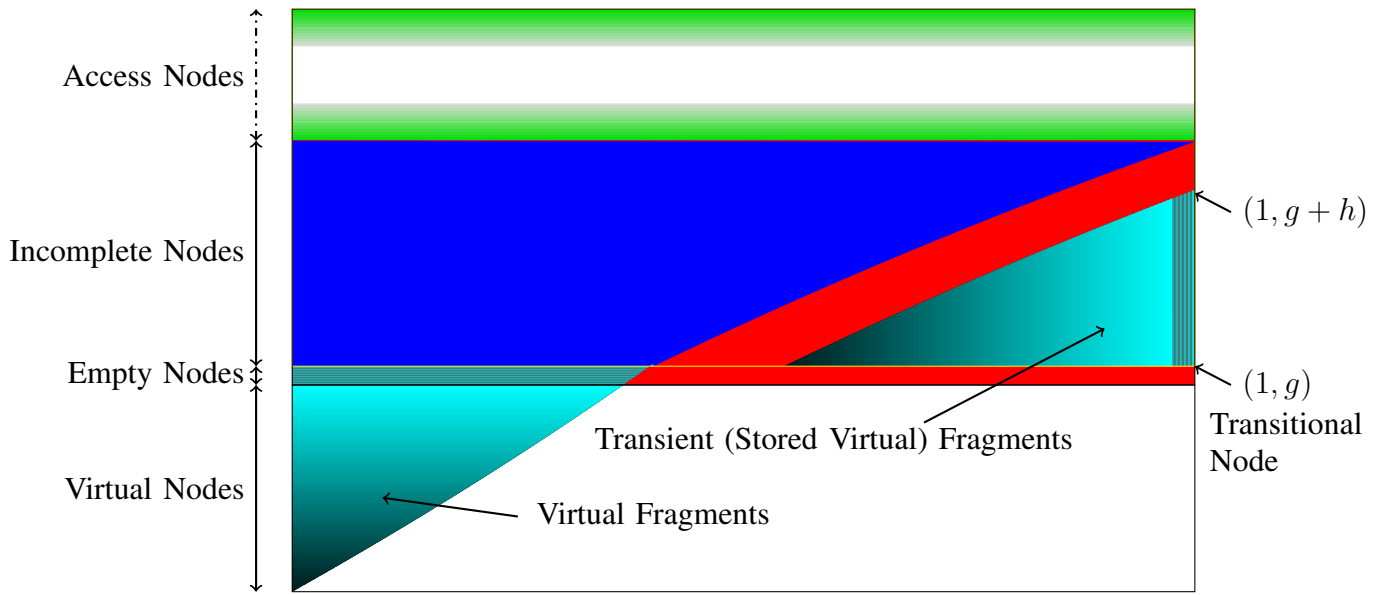
\begin{figure}
\begin{tikzpicture}[x=12cm,y=5cm]

\draw[fill, top color=white, bottom color=green,draw=none] (0,\ANb) rectangle (1,\ANb+0.1);
\draw[fill, top color=green, bottom color=white,draw=none] (0,0.9) rectangle (1,1);
\draw[fill=red,draw=none] (0,0) rectangle (1,\ANb);
%   \fill[red, domain=0.4:1, variable=\x,samples=1000]   (0.40,0.0) -- plot ({\x},{(1-exp(-0.58*(\x-0.4))})   --  (1,0) --cycle;
   \fill[blue, domain=0:1, variable=\x,samples=1000]   (0,\gh) -- plot ({\x},{ max(\gh,\gh+2.23*(1-exp(-0.52*(\x-0.4)) }) -- (0,\ANb) -- cycle;
    \fill[domain=0:0.4,top color = cyan,bottom color = cyan!10!black,  variable=\x,samples=1000,opacity=0.5]   (0,\gh) -- plot ({\x},{\VNb+\gh - 2.6*ln((1.0-0.52*\x)) }) -- (0.4,0+\gh) -- cycle;
    \fill[domain=0:0.4,top color = cyan,bottom color = cyan!10!black,  variable=\x,samples=1000,opacity=1]   (0,0) -- plot ({\x},{min(0,\VNb+\gh - 2.6* ln((1.0-0.52*\x)) ) }) -- (0.4,0) -- cycle;
   \fill[domain=1.0-0.455:1.0, left color = cyan!10!black,right color = cyan, variable=\x,samples=1000,opacity=0.5]   (1.0-0.455,\gh) --plot ({\x},{\gh+2.23*(1-exp(-0.52*(\x-(1.0-0.455)))}) -- (1.0,\gh) -- cycle;
   \fill[domain=1.0-0.455:1.0-0.025, left color = cyan!10!black,right color = cyan, variable=\x,samples=1000,opacity=1.0]   (1.0-0.455,\gh) --plot ({\x},{\gh+2.23*(1-exp(-0.52*(\x-(1.0-0.455)))}) -- (1.0-0.025,\gh) -- cycle;
%\node[below] (v11a) at (1+\nudge,0.0) {};
%\node[above] (v11b) at (1+\nudge,0.025) {};
%\draw[thick,<->]  (v11b) -- (v11a) node[midway,right] { $(1,g)$ };
\draw[thick, ->] (1+\nudge,0.0) --  (1.0,\gh)   node[at start,right]  { $(1,g)$ };
\draw    (1.1,-0.15)  node [text width=2cm] { Transitional Node };
\draw[thick, ->] (1+\nudge,0.0+0.46) --  (1.0,\gh+0.46)   node[at start,right]  { $(1,g+h)$ };

\draw[color=yellow] (0,\gh) rectangle (1,1);
\draw (0,0) rectangle (1,1);
\draw (0,\VNb+\gh) rectangle (1,0);

\draw  (0.4,-0.35) node {Virtual Fragments};
\draw[thick, ->] (0.25,-0.35) --  (0.1,-0.3);

\draw  (0.6,-0.15) node { Transient (Stored Virtual) Fragments};
\draw[thick, ->] (0.7,-0.1) --  (0.9,0.15);
  
\node[below] (ndVNb) at (-\nudge,\VNb+\gh) {};
\node[above] (ndVNt) at (-\nudge,0.0) {};
\node[below] (ndENb) at (-\nudge,0.0) {};
\node[above] (ndENt) at (-\nudge,\gh) {};
\node[below] (ndTNb) at (-\nudge,\gh) {};
\node[above] (ndTNt) at (-\nudge,\ANb) {};
\node[below] (ndANb) at (-\nudge,\ANb) {};
\node[above] (ndANt) at (-\nudge,1) {};
\draw[thick,<->]  (ndVNb) -- (ndVNt) node[midway,left] { Virtual Nodes };
\draw[thick,<->]  (ndTNb) -- (ndTNt) node[midway,left] { Incomplete Nodes };
\draw[thick,<->]  (ndENb) -- (ndENt) node[midway,left] { Empty Nodes };
\draw[thick,<->, dash dot]  (ndANb) -- (ndANt) node[midway,left] { Access Nodes };
\end{tikzpicture}
\caption 
{\label{fig:basicrepack}
Visualization of partially virtualized repair queue.  The virtual fragments are stored on the transient nodes as indicated.
The boundary between the transient nodes and the empty nodes (the yellow line) depicts the transitional  node.
%Here we chose parameters $\nlr \Ttot = 0.58$ and $\beta'' = 0.263$ yielding $z = 0.4.$
%Setting $\nvfobj = 0.58^{-1}$ we obtain 
}
\end{figure}

As described above, in typical operation the basic liquid repair system leaves approximately half of the available storage overhead unused and
this deficiency accounts for its suboptimal repair read rates.
One possible remedy for this deficiency is to virtualize the storage of the bottom portion of the repair queue by placing those fragments in the remaining unused portion of the repair queue, see Fig. \ref{fig:basicrepack}.
This is the approach considered in this section.
The repacking perturbs the dynamics of the repair process and the typical form of the incomplete repair queue changes.
It turns out that under an appropriate form of the repacking the repacked portion of the queue (asymptotically) exactly fits in the remaining unused portion and, consequently, optimal read repair rates with full storage utilization can be asymptotically achieved.
%We will consider several variants of the partially virtualized scheme in which we may allow variation in the length of the code and even in the number of nodes in the system.

In the basic liquid system the storage overhead $\beta$ corresponds with the code rate in that $\kcode = (1-\beta) N$ and $r = \beta N.$
In the partially virtualized scheme we will instead use a larger codelength to accomodate virtualized nodes.
When a storage node fails, its physical replacement will assume the identity of a virtualized node including adopting its associated code symbol. 
The symbol associated to the failed node will instead be assigned to a new virtual node which is added to the system.

The number of fragments regenerated per object repair will vary slightly from repair to repair but it will approximate $ (\beta + \beta_v) N$  where $\beta_v N$  represents a number of virtualized nodes.
The partially virtualized scheme is roughly similar to a basic liquid system with code rate $\frac{1-\beta}{1+\beta_v}$ and $(1+\beta_v) N$ nodes.
Consider in such a  basic liquid system 
the portion of the repair queue comprising the bottom $\beta_v N$ nodes.
The number of fragments stored in those nodes is a non-decreasing function of the their height ($y$ position) in the repair queue.
Let $x_v (t)$ denote the largest $x$ position of a fragment for the node in node position $\beta_v N - 1.$
There are therefore $x_v (t) \nobject$ objects that store fragments in this lower portion of the repair queue.
In the partially virtualized scheme the bottom $\beta_v N$ nodes are virtual, i.e., they do not exist as actual nodes.
The fragments that appear in these nodes in the repair queue are virtual and they are actually stored in an otherwise unused portion of the actual repair queue.
The thus-stored virtual fragments will be referred to as {\em transient} fragments and, for transient fragments, the code symbol associated to the fragment is not the one associated to the node on which they are stored.
As part of the repair process transient fragments are later moved (copied) to their intended final location on the node associated to their code symbol.
Fragments that are written to their final location will be called {\em settled} fragments.
Thus, when a transient fragment is copied to its intended location it becomes a settled fragment.

\subsection{The Repacking Scheme}

We will now give a more detailed and formal description of the partially virtualized  transient fragment packing scheme,
which is depicted in Fig. \ref{fig:basicrepack}.
In the rough analogy to the basic liquid system made above we assumed $\beta_v N$ virtual nodes and $N$ actual nodes.
The repair process proceeds much as in the basic liquid case from the perspective of the virtual queue, but in the actual system this entails
the writing and moving transient fragments and the promotion of virtual nodes to actual nodes.
We assume that at any time that there is one node (at most) in transition between the virtual state and the actualized state. 
We refer to this node as the {\em transitional node}.
Transient fragments that are regenerated during the repair process are written exclusively to the transitional node.
For purposes of system analysis we will treat the transitional node as virtual until the point in time where its initial transient
and settled fragments have been completely written, at which point
we say that the node is {\em launched}.
We will discuss later how the writing of the transitional node can be additionally protected so that
this assumption could be supported in practice, but
its main purpose is to avoid analytical complications inherent in the potential failure of the transitional node.
In practice, such a failure would be no more damaging then the failure of any another node.

At any time $t$ the nodes in the top $(1 - g_v(t)) N$ positions are actual (post launch) and the transitional node will be in node position $g_v(t)N-1.$
The transitional node must be physically real for (standard) repair to proceed so
 if $g_v(t)N = 0$ then standard repair is suspended.
(Later we will add an ancillary repair process that can proceed when $g_v(t)N = 0.$)
Assuming standard repair proceeds,  a transitional node eventually {\em completes}, is launched, and ceases to be the designated transitional node.
At that time the node immediately below in the queue becomes the transitional node and $g_v(t) N$ is decremented by $1.$
Any actual nodes below the transitional node, i.e. those in node positions $[0 : g_v(t)N-1)$ can be considered as physically empty
although we depict them as holding virtual fragments.  This implies, for the purposes of our model, that they cannot fail.
The virtual fragments associated to those nodes are stored as transient fragments spread
across those nodes in node positions $[g_v(t) N : (g_v(t)  + h_v(t))N)$ where we have introduce $h_v(t) N$
to represent the number of launched transient fragment carrying nodes.
The repacking scheme uses $h_v(t) N$ consecutive nodes in the actual repair queue to store the transient fragments
and these nodes occupy node positions $[g_v(t)N : g_v(t)N+h_v(t)N).$

We introduce a dimensionless paramater $\nvfobj$ where those transient fragments associated to $\nvfobj \frac{\nobject}{N}$ consecutive objects
will be stored on a single node.   Correspondingly, the transitional node remains transitional through $\nvfobj \frac{\nobject}{N}$ object repairs,
at which point it ceases to be the transitional node and is launched as an actual node.
The time required to repair  $\nvfobj \frac{\nobject}{N}$ objects under continuous standard repair ($g_v(t) > 0$) at rate given by $\Tsysrep$ will be denoted by
\[
\Tnode := \frac{\nvfobj}{N} \Tsysrep\,.
\]
Hence, under constant rate continuous standard repair, a transitional node is completed and launches as an actual node every $\Tnode$ time units.

If a node that carries transient fragments fails, a time $t$ say, then those transient fragments are erased. 
The corresponding  $\nvfobj \frac{\nobject}{N}$ contiguous objects in the repair queue lose their virtual fragments and
the number of intact fragments for those objects is immediately reduced to $(1-g_v(t-))N.$
We refer to this as a {\em  virtual fragment loss event}.
Upon a virtual fragment loss event we move the affected $\nvfobj \frac{\nobject}{N}$ objects forward in the repair queue, placing them directly in front of those objects that still have virtualized fragments.
Thus, these objects are moved to those consecutive positions ending at $x_v(t-),$ and $x_v(t+)$ is correspondingly reduced.
While this reordering disrupts the invariant cyclic repair ordering of the basic liquid system, it preserves the fragment symbol ordering property of the repair queue,
i.e., after the reordering the set of intact fragment code symbols for objects is ordered by inclusion along the queue.

%The dynamics of the virtual portion of the repair queue can be described in both the virtual portion of the queue and in its image as stored in the actual queue.
%We will use both descriptions to facilitate a better understanding of the mapping between the virtual nodes and their image.
To give a more precise description of the partial virtualization repacking scheme and to facilitate analysis
 we will assign specific repair queue positions to the transient fragments.
We extend the coordinates of the basic liquid system so that the purely virtual node portion of the queue is represented by $y<0.$
The physical nodes and their storage is represented by the area $(x,y) \in [0,1)^2$ just as in the basic liquid case,
(although the launched nodes are represented by the area $(x,y) \in [0,1)\times [g_v(t),1).$)
We extend the definition of $f(x,t)$ to be consistent with this,
i.e., $(1-f(x,t))N$ represents the number of intact fragments for an object in position $x$ at time $t.$
For $x \le x_v (t)$ we therefore have $f(x,t) \le 0.$ 
%and $(g_v(t) N+1-f(x,t) N)$ represents the number of virtual fragments for the associated object.

The portion of the repair queue consisting of the settled fragments for objects in $x$-positions with $x > x_v (t),$  i.e., those without virtual fragments, 
behaves nearly identically to a reduced size basic liquid repair queue.
While the basic liquid repair queue has a fixed number of objects and nodes, the non-virtualized portion of the repair queue has a fluctuating number of objects and nodes because of fluctuations in $x_v (t)$ and $g_v(t)$ respectively.
For large systems, however, $x_v (t)$ and $g_v(t)$ will concentrate around their expected values.
Another difference with the basic liquid system is the presence of the transient fragments, i.e., the image of the virtualized fragments.
The presence of the transient fragments has, however, no direct effect on the dynamics of the non-virtualized portion of the queue as long
as there is sufficient capacity on each actual node to store both sets of fragments.

At the granularity of fragments it is natural to understand each settled and virtual fragment as occupying a storage area in the repair queue corresponding to a rectangle of size $\frac{1}{N} \times \frac{1}{\nobject}.$
To specifiy positions for transient fragments and associate them to their corresponding virtual fragment positions we introduce a linear (for fixed $t$) map $\varphi_t: \reals^2 \rightarrow \reals^2$   
defined by $\varphi_t (x,y) = (1-\nvfobj (y-g_v(t)) , g_v(t) + (1+\nvfobj^{-1}  x )).$
Note that $\varphi_t$ is area preserving, that $\varphi_t(x+\frac{\nvfobj}{N},y) =\varphi_t(x,y) + (0,\frac{1}{N}),$
and that $\varphi_t(x,y+\frac{1}{N}) =\varphi_t(x,y) + (\frac{\nvfobj}{N},0).$
Under the mapping $\varphi_t$ the $\frac{1}{N} \times \frac{1}{\nobject}$ rectangle associated to a virtual fragment
is a rectangle of the same area but with size
%\(
%\frac{1}{N(\nvfobj \frac{\nobject}{N})} \times \frac{(\nvfobj \frac{\nobject}{N})}{\nobject}
%=
\(
\frac{1}{\nvfobj \nobject } \times \frac{\nvfobj }{N}\,.
\)
We can conceive of the associated transient fragment as occupying this rectangle.
While virtual fragment rectangles are packed horizontally along node strips the transient fragments appear in stacks of height $\kappa$ on a node.
Thus, if we consider the virtual fragments for objects in $x$ position $\frac{\nvfobj}{N} [k,k+1)$  virtually stored on a virtual node with associated strip $\frac{1}{N}[j,j+1),$  ($j<0),$
then these fragments occupy a virtual rectangular area of size  $\frac{\nvfobj}{N} \times \frac{1}{N}$ and
under the mapping $\varphi_t$ their image occupies a rectangle of precisely the same size on a single node.
At the granularity of fragments, the mapping can be understood this way.
We note that the $x$ position of a transient fragment has no physical significance, since it is only necessary that the fragment be stored on the corresponding node, and we allow the $x$ position to change with time.
The $y$ location of transient fragments, however, indicates the node on which the object is stored and the $y$ positions change correspondingly.
See Fig. 

%\marginpar{NEEDS REVIEW}
%For the analysis of MTTDL it will be convenient to consider these nodes as uneraseable.  We will present a system justification for this in which the data on those nodes is duplicated so
%that, when one fails, it can be reproduced by copying.  In a real implementation the duplication is not guaranteed to be infallible, but if duplicate copies of a fragment are both lost
%the consequence will simply be a temporarily less efficient repair event and not a data loss event.  

%We now extend the definition of $h_v(t)$ to $h_v(x,t)$ so that $h_v(x,t) N$ denotes the number of nodes carrying transient fragments (excluding the transitional  node) in position $x.$   We therefore have $h_v(t) = h_v(1,t)$ and we will continue to use the earlier notation where convenient,
%hopefully without causing confusion.

\subsection{Node Failure and Object Repair}

We now describe the failure and repair processes in greater detail.  

\subsubsection{Standard object repair}
Standard object repair can proceed only when $g_v(t) > 0$ so that the transitional node exists as a physical node.
%We assume that for each object repair a constant $\mtrans$ number of virtual fragments are generated and written as transient fragments to the transitional node.
The object under repair at time $t$ has $(f(1,t)-g_v(t)) N$ fragments erased from among the top $(1-g_v(t)) N$ nodes in the repair queue.
During repair, these fragments are regenerated for the object and written to those nodes as settled fragments just as in the basic liquid system.
%which is in $y$ position $g_v(t)-\frac{1}{N}.$ 
One further settled fragment is regenerated for the transitional node.
Note that this fragment may be viewed as both settled or transient, since as a 'transient' fragment it would later be copied to the transitional node to become a 'settled' fragment.
In addition, a constant $\beta_v N$ number of virtual fragments are generated and written as transient fragments to the transitional node.
These transient fragments correspond to virtual fragments on virtual nodes appearing in the repair queue below the transitional node.
Thus, a total of $(f(1,t)-g_v(t)) N + 1 + \beta_v N$ fragments are regenerated. See Fig. \ref{fig:basicrepack}.

\subsubsection{Atomic nature of Transitional node writing}

Assuming that the transitional  node cannot fail is equivalent to treating the writing of the transitional  node as an atomic event.
In an actual implementation a failure of the  transitional  node could be handled by forwarding in the repair queue those among the $\nvfobj \frac{\nobject}{N}$ objects whose repair had been completed.
Alternatively, if a copy of those fragments had been maintained elsewhere then the transitional  node could be reconstituted on a replacement node.
The transient fragments copied to realize virtual fragments would still be available since the memory occupied by those nodes is not released until the transitional  node
is complete, so those fragments would be available for a replacement transitional  node.
In any case, the failure of a transitional  node is a small perturbation in the repair process and could easily be absorbed in any practical implementation without significantly affecting the long term behavior of the repair process.
When a transient fragment carrying node fails we view its transient fragments as lost.  It is possible, however, that some of those fragments
had already been copied to the transitional node.   For simplicity we will ignore this possibility and treat those fragments as completely lost.  This is equivalent
to assuming that the copying process happens instantaneously upon completion of the transitional node.

\subsubsection{Transitional node completion and actual node launch}
A particular transitional node is used to store transient fragments regenerated from $\nvfobj \frac{\nobject}{N}$ successive object repairs.
Once those repairs are complete that node will store $\beta_v N  \nvfobj  \frac{\nobject}{N} = \beta_v  \nvfobj  {\nobject}$ transient fragments
and we say that the transitional  node has been completed.
When a transitional node \node~completes it becomes an actual (transient fragment carrying) node with node position $g_v N.$
At this time the node fully enters the system and it's lifetime properly begins, hence
we refer to this as the node's  launch.
For a node $\node$ we will use $\launchtime(\node)$ to denote its time of launch.
The node will fail some time later, denoted $\failtime(\node)$ and $\failtime(\node) - \launchtime(\node)$ is an independent exponentially distributed random variable with rate $\lambda.$
Note that actual nodes are ordered in the repair queue by launch sequence.  We will introduce the notation $\launchid(\node)$ to denote the 
relative order of the launched nodes.  Initially ($t=0$) all actual nodes will possess a value for $\launchid(\node) \le 0$ and they will be correspondingly ordered in the repair queue.
Subsequent launches (for $t > 0$) will be indexed starting from $\launchid(\node) = 1.$

%It may hold fewer because some fragments may have been removed due to devirtualization.
Upon completion the  transitional node ceases to be the transitional node and $g_v  N$ reduces by $1.$ 
More precisely, we have $g_v(\launchtime(\node)+)  N = g_v(\launchtime(\node)-)N-1$
and the node position of $\node$ at time $\launchtime(\node)+$ is $g_v(\launchtime(\node)+)  N.$
The node immediately below, in node position  $g_v(\launchtime(\node)+)  N - 1 =  g_v(\launchtime(\node)-)N-2,$ assuming it exists as a physical node, now becomes the transitional node.
(We will ignore the possibility of simultaneous node failure and node launch.)
While the $\nvfobj \frac{\nobject}{N}$ object repairs associated to one transitional node are proceeding, the virtual fragments residing on the transitional nodes are realized, i.e., made settled, by copying the transient images of those fragments to the transitional  node.
Completion of the  transitional  node also entails the completion of this copy process; we assume it is accomplished by the time the last object repair
for that transitional  node is completed.
Treating the completion as an atomic event, the number of transient fragments copied to the transitional node is given by $h_v(\launchtime(\node)-) N \nvfobj \frac{\nobject}{N} = h_v(\launchtime(\node)-)  \nvfobj {\nobject}.$
Note that the number of initial settled fragments on the node upon launch is slightly larger, being given by $(h_v(\launchtime(\node)-) N + 1 )\nvfobj \frac{\nobject}{N} = (h_v(\launchtime(\node)-)+\frac{1}{N}) \nvfobj {\nobject}.$
Upon transient node completion the storage capacity associated to copied transient fragments is released, i.e., it is made available for overwriting.  
Conceptually, in the context of the repair queue, we can view those transient fragments as having been erased.
By the definition of $\varphi_t,$ the decrementing of $g_v N$ at completion times corresponds to increasing the $x$ positions of the remaining transient fragments by 
$\frac{\kappa}{N}$ (see Fig. \ref{fig:basicrepack}).

For $t \ge \launchtime(\node)$ the settled fragments on node $\node$ occupy x-positions $[0,z_s(\node,t)]$ where we have introduce the 
parameter $z_s(\node,t)$ to indicate the right extreme position of the settled fragments.
Hence $z_s(\node,\launchtime(\node)+) = \kappa  (h_v(\launchtime(\node)-)+\frac{1}{N}) .$
Similarly, for $t \ge \launchtime(\node)$ the transient fragments on $\node$ occupy x-positions $[z_v(\node,t),1]$ where we have introduce the 
parameter $z_v(\node,t).$
At launch, the number of transient fragments on the node is $\beta_v N \kappa \frac{\nobject}{N} = \beta_v  \kappa \nobject,$ 
hence $z_v(\node,\launchtime(\node)+) = 1- \beta_v  \kappa.$
At times of susbsequent transient node completions $z_v(\node,t)$ increments by $\frac{\kappa}{N}$ until $z_v(\node,t)$ reaches $1,$
which occurs after $\beta_v N$ such launches.
If a launched node $\node$ has no transient fragments stored on it at time $t$ then we say $z_v(\node,t) = 1\!+.$
Let us denote the time that $z_s(\node,t)$ reaches $1$ as $\settletime(\node).$
Note that both $z_v(\node,t)$ and $z_s(\node,t)$ can be defined and uniquely determined even if $\node$ has failed (they take the value they would have
if some other node had failed instead) and we occasionally use this fact.

Note that we generally require $z_s(\node,t) \le z_v(\node,t)$ which limits the choice of $\beta_v.$
In particular this requires $ \kappa  (h_v(\launchtime(\node)-)+\frac{1}{N}) \le 1- \beta_v  \kappa.$
Should this inequality be violated at a launch time for some node $\node,$  then let us stipulate that the right-most settled fragments targeted for $\node$ will be dropped to reduce $z_s(\node,t).$
In the analysis below we will choose system parameters so that this event is exponentially rare in $N.$
As standard repair proceeds the two values $z_s(\node,t)$ and $z_v(\node,t)$ increase essentially in lock step.  If we treat transitional node completion as
an atomic discrete event then the two positions increase exactly together by $\frac{\kappa}{N}$ upon each transitional node completion.
If we view the generation of settled fragments at the granularity of objects then the settled fragments increase with finer granularity, 
but the two are re-synchronized upon transitional node completion.
More precisely, $z_s(\node,t)$ increases by $\frac{1}{\nobject}$ in $\frac{\Tsysrep}{\nobject}$ intervals whereas
$z_v(\node,t)$ increases by $\frac{\kappa}{N}$ in intervals of length ${\Tnode} = \frac{\kappa}{N}\Tsysrep.$
Thus, assuming an initial negligible margin sufficient to store $\kappa \frac{\nobject}{N}$ fragments, there is always sufficient capacity on the node to store its assigned transient fragments.
For convenience we will generally adopt the atomic event viewpoint.

\subsubsection{Ancillary Repair}
When $g_v(t) = 0$ standard repair cannot proceed.  The system could, however, continue to repair objects without generating transient fragments.
More specifically, the {\em ancillary repair process}  regenerates those $f(1,t) N$ fragments missing from actual nodes and writes them to the corresponding actual nodes.
Thus, the repaired object will have $N$ intact fragments after repair.  It can then be place in the repair queue in position $x_v(t),$ immediately in front of
those objects that possess virtual fragments.  This process resembles basic liquid repair for the subsystem consisting of those objects in $x$-positions greater than
$x_v(t).$  Note that this involves no change in the set of transient fragments.

Under ancillary repair the value of $z_s(\node,t)$ increases for the bottom $f(1,t) N$ nodes while $z_v(\node,t)$ remains unchanged.
If for some such node it then arises that $z_s(\node,t) = z_v(\node,t)$ then there is no remaining room to write the regenerated fragments on that node.
Let us stipulate, in that case, that the settled fragment corresponding to the object with maximal $x$ position among objects possessing settled fragments on 
$\node$ will be overwritten by the regenerated fragment.  
With this stipulation the structure of the repair queue remains intact and we continue to have $z_s(\node,t) = z_v(\node,t).$

\subsubsection{Transient and Settled Fragment Processes}
At any time $t$ the number of launched nodes storing transient fragments is $h_v(t) N$ and
the number of transient fragments on each node is an integer multiple of $\kappa \frac{\nobject}{N}.$
At transitional node completion times each transient fragment carrying node releases the storage occupied by the $\kappa \frac{\nobject}{N}$ fragments that were copied to the transitional node.
In addition, a new transient fragment carrying node is launched carrying $\beta_v \kappa \nobject$ transient fragments.
The following result, which we state without further proof, captures the basic dependence.
\begin{lemma}
Let $\launchid(\node) \le \launchid(\node')$ and assume at time $t$ that node $\node$ stores transient fragments.
Then $z_\mtrans(\node,t) - z_\mtrans(\node',t)  =  (\launchid(\node')-\launchid(\node)) \frac{\kappa}{N}.$
\end{lemma}

When a transient fragment carrying node fails $h_v(t)N$ reduces by $1.$
At a transitional node completion time the above Lemma implies that $z_\mtrans(\node,t)$ may reach $1$ for at most $1$ transient fragment carrying node.
Thus $h_v(t)N$ will either remain unchanged or increase by $1.$  Transitional node completion and transient carrying node failure are the only two events
that modify the transient fragments.

\begin{lemma}[Montonicity]\label{lem:monotonicity}
Let $\node$ and $\node'$ be successively launched nodes, i.e. $\launchid(\node) = \launchid(\node') - 1,$ then assuming neither node has failed
and $t \ge \launchtime(\node')$, we have
\begin{align*}
z_s(\node',\launchtime(\node')+)- z_s(\node,\launchtime(\node)+)   & \le  \frac{\kappa}{N} \\
 z_s(\node,t) - z_s(\node',t)  & \ge 0
\end{align*}
\end{lemma}
\begin{IEEEproof} 
Consider the first inequality.  If $z_s(\node,\launchtime(\node)+) = 1-\beta_v  \kappa$ then the result is immediate, so we assume
$z_s(\node,\launchtime(\node)+) = \frac{\kappa}{N}(h_v(\launchtime(\node)-)N + 1) < 1- \beta_v\kappa.$
Now, as described above $h_v(t) N$ can increment (by $1$) only upon completion of a transitional node and can otherwise only decrease. 
Hence  $\frac{\kappa}{N}(h_v(\launchtime(\node')-)N + 1) \le \frac{\kappa}{N}(h_v(\launchtime(\node)-)N + 2)$ which now gives the first result.

Between the launch of $\node$ and the launch of $\node'$ exactly $\kappa\frac{\nobject}{N}$ objects are repaired under standard repair.
Hence $z_s(\node,\launchtime(\node')+) \ge  z_s(\node,\launchtime(\node)+)+ \frac{\kappa}{N} \ge z_s(\node',\launchtime(\node')+).$ (Note that the first inequality is an equality if no ancillary repair has occurred in the meantime.)
For $t \in (\launchtime(\node'),\settletime(\node)]$ the quantity $z_s(\node,t)-z_s(\node',t)$ is unchanging unless either $z_s(\node,t)=z_v(\node,t)$ or $z_s(\node',t)=z_v(\node',t).$
If $z_s(\node,t)=z_v(\node,t)$ then we have $z_s(\node',t)\le z_v(\node',t) <z_v(\node,t).$
If it first occurs that $z_s(\node',t)=z_v(\node',t)$ then, after that we have $z_s(\node,t) \in [z_v(\node',t),z_v(\node,t)].$
Thus, the second inequality holds in all cases.
\end{IEEEproof}

\subsubsection{The Node Launch Position Process}

The value of $g_v(t)N$ increases by $1$ when a launched node fails.
The value of $g_v(t)N$ decreases by $1$ upon a transitional node completion.
The rate of launched node failure is given by $\lambda (1-g_v(t)) N.$
Thus, under our current assumption the integer valued $g_v(t)N$ process follows closely the queue size of a machine interference problem (MIP) \cite{saaty1961elements,MIPSA1985}.
If we assume  constant rate  standard repair then $g_v(t)N$ is precisely an MIP with constant service time.
In Kendall notation this is an $M/D/1//N$ queing problem.  In the MIP problem it is assumed that there are $N$ machines whose
time to failure is an exponentially distributioned random variable with rate $\lambda.$  Upon failure the machines enter a queue for service and the repair time
is another random variable, in our case assumed to be a deterministic constant.  In some MIP models one assumes a finite capacity for the queue, but in our case
this is immaterial.

More loosely, the $g_v(t)N$ process behaves much like the number of users in a single server queue. 
The service time of the queue is a constant under the assumption of fixed repair rate.
A system designer, however, would be free to vary the repair rate as a function of the state of the system.
The number of users in the queue cannot be arbitrarily  large since having more than $N-\kcode$ users in the queue implies data loss.

\subsubsection{Survivor process}

For a launched node \node~and $t\ge \launchtime(\node)$ let $\survive(\node;t)$ denote the number of nodes
launched since $\launchtime(\node)-$ (including \node) that have not failed by time $t.$
Thus, formally
\[
\survive(\node;t) =  \{ \node'  :  \launchid(\node') \ge \launchid(\node),\failtime (\node') > t \} 
\]
where we recall $\failtime (\node)$ denotes the failure time of node $\node.$

%$|\survive(\node;t)|+ g_v(t)$ is closely related to $f(z_s(\node,t) ,t) .$
\begin{lemma}
For $\launchtime(\node) \le t < \settletime(\node)$ we have
\[
f(z_s(\node,t)- ,t) \le |\survive(\node;t)| + g_v(t) \le f(z_s(\node,t)+ ,t)
\]
\end{lemma}
\begin{IEEEproof}
By Lemma \ref{lem:monotonicity}, for any node $\node' \in \survive(\node;t)$ we have
$z_s(\node',t) \le z_s(\node,t).$ 
It follows that $f(z,t)N \ge |\survive(\node;t)| + g_v(t)$ for any $z > z_s(\node,t).$

For any node $\node''$ with $\launchid(\node'') \le \launchid(\node)$ we have $z_s(\node'',t) \ge z_s(\node,t).$
Hence, assuming at least one such node survives at time $t,$ we have $f(z,t)N \le |\survive(\node;t)| + g_v(t)$ for $z < z_s(\node,t).$ 
If no such node survives then $g_v(t)N + |\survive(\node;t)| = N.$ 
\end{IEEEproof}

We define a data loss event as a node loss that results in at least one object having fewer than $\kcode$ intact fragments.
(For purposes of analysis of system dynamics we generally assume that the repair proceeds regardless of this event.)
Suppose a first data loss event occurs at some time $t.$  This implies that a fully settled node failed at time $t$ and the node $\node$ that then enters node position
$N-\kcode$ is unsettled, i.e., $\settletime(\node) > t.$   It follows that $|\survive (\node;t+)|+ g_v(t+) = N-\kcode$ and since $|\survive (\node;t)| + g_v(t)$ can only increase with $t$ we have
$|\survive (\node;\settletime(\node))| + g_v(\settletime(\node)) \ge N-\kcode.$  
Since $\settletime(\node) - \launchtime(\node) \le \Tsysrep$ this yields the following result.
\begin{lemma} 
If $|\survive (\node;\settletime(\node))|+ g_v( \settletime(\node) ) < N-\kcode$ for all nodes with $\settletime(\node) \le t$ then $f(1,s)N < N-\kcode$ for all $s \in [0,t - \Tsysrep].$
\end{lemma}
%\begin{IEEEproof}
%\end{IEEEproof}

%Note that we also have almost everywhere the bound
%\[
%|\survive(\node;t)| \le  \Big\lceil \frac{t - \launchtime(N) }{\Tnode} \Big\rceil
%\]
%For $t - \launchtime(\node) \in [0,  \Mv \Tnode]$ we have $h_v (z_\mtrans(\node,t),t) N = \survive(\node;t).$

\subsection{Analysis of Continuous Standard Repair}

In this section we present an analysis of a system under continuous standard repair at constant repair rate.
We set the system parameters so that $g_v(t) N$ gravitates towards a value  $\gstar N$ for a small positive design parameter $\gstar.$
We consider what is essentially a single busy period of continuous standard repair and show that its expected length (without data loss) is exponential in $N.$
In this mode of operation the storage system always has a non-zero number of empty physical nodes. 
In practice such a form of operation could be viable if there are many such systems running on a shared set of storage nodes, so that the
unused nodes could be aggregated across a much larger system and their number kept relatively small.
For a single such system a more practical version would likely use faster repair with occasionaly stanard repair suspension.
We will show that in the large system limit ($N \rightarrow \infty)$ that with properly chosen parameters we can have 
\begin{align*}
\text{MTTDL} & \rightarrow \infty \\
\text{repair read rate} & \rightarrow (\lambda \Tsysrep)^{-1}(\beta- \ln(1-\beta))
\end{align*}
We conjecture that the factor $\beta- \ln(1-\beta)$ is optimal.

Long-lasting continuous standard repair must balance the node production rate with the node loss rate. 
We will choose $\gstar$ as a (small) positive target value for $g_v(t).$ 
When $g_v(t) = \gstar$ the node production rate and the node loss rate will be equal, which results in
\[
\nvfobj = ((1-\gstar) \lambda \Tsysrep)^{-1}.
\]
A transitional node completes every $\Delta_t = \kappa \frac{\Tsysrep}{N}$ time units, hence nodes are launched at rate $1/\Delta_t$ while they fail at rate
$\lambda (1-g_v(t))N.$  Note that we have the relation
\[
\lambda \Delta_t = \bgstar N\,
\]
where $\bgstar = 1-\gstar.$

After a node is launched $\beta_v N$ subsequent transitional  node completions are required to clear out the transient fragments
and this will occur, under continuous standard repair, after an elapsed time $\beta_v N \Tnode = \beta_v \kappa \Tsysrep.$
Similarly,  under continuous standard repair, we have $\settletime(\node)-\launchtime(\node)= (1-z_s(\node,\launchtime(\node)) {\Tsysrep}.$

\subsubsection{Initial Condition}

We aim to show extremely long operation of the system, but to be concrete we introduce an initial condition set essentially to the expected behavior,
 in particular we assume  $g_v(0)  =  \gstar.$
We construct the initial condition ($t=0+$)  by supposing that nodes had been launched at times $0,-\Tnode,-2\Tnode,\ldots.$
Some of these nodes will be assumed to have failed by time $t=0.$
Nominally, a  node launched at time $-k\Tnode$ would have survived to time $t=0$ with probability $e^{-\lambda \Tnode k}.$
Among nodes launched at times $0,-\Tnode,-2\Tnode,\ldots,-d\Tnode$ the expected number of nodes surviving at time $t=0$ is given by
\begin{equation}\label{eqn:initialsurvive}
S_d := \sum_{k=0}^d e^{-\lambda \Tnode k} = \frac{1-e^{-\lambda \Tnode (d+1)}}{1-e^{-\lambda\Tnode}}.
\end{equation}
We will consider that the node launched at time $-k\Tnode$ has failed by time $t=0$ if $\lceil S_{k-1} \rceil = \lceil S_k \rceil.$
This implies that the node launched at time $0$ still survives ($S_0 = 1$).
The number of nodes surviving at time $0$ from $-d\Tnode,...,0$ is $\lceil S_d \rceil.$
The node $\node$ in node-position $\gstar N + j$ at $t=0+$ has $-\launchid(\node)$ equal to the smallest $d$ such that $S_d > j.$
Note that $S_\infty > \bgstar N$ so all initial operational nodes are assigned a launch index.
It follows that for launched node in position $\gstar N + j$ at $t=0+$ we have
\(
-\launchid(\node) = \lfloor -(\lambda\Tnode)^{-1} \ln (1-(1-e^{-\lambda\Tnode}) j)  \rfloor \,.
\)

For all launched nodes we assume $z_s(\node,0+) = z_v(\node,0+),$ where $z_v(\node,0+) = (1-\beta_v\kappa) - \launchid(\node)\Tnode.$
We consider the initial nodes in node position $\gstar N + j$  for $j < 0,$ to have not been launched.

%Then  $\xsettled(\node,0) = Z_v.$
%Define a sequence $k_1,k_2,...$ as 
%\[
%k_i = \argmin_k \frac{1-\delta_{k_i+1} }{1-\delta} - \sum_{i'=1}^{i-1} \delta^{k_i} \le k_i - \frac{1}{2}
%\]
%Then for $j=1,2,...$ we set $\xsettled(\node,j) = \xsettled(\node,j-1) + \frac{\kappa}{N}$
%unless $j= k_i$ in which case we skip.
%
%OR use asymptotic curve
%\[
%\xsettled(\node_j,0) = Z_v - 
%\frac{\kappa}{N} \lfloor
%-\frac{N}{\kappa \lambda T} \ln(1 - j (e^{\frac{1}{\bgstar N}} - 1))
% \rfloor
%\]
%based on $\frac{1-e^{-\lambda T Z}}{e^{\frac{1}{\bgstar N}} - 1} h(Z) N$
%until $\xsettled(\node_j,0) > 1.$
%
%Set $L(\node_j) = -j.$ (But then those with $j=k_i$ should be deleted....)
%or 
%\(
%L(\node_j) =  -\lfloor
%-\bgstar N \ln(1 - j (e^{\frac{1}{\bgstar N}} - 1))
% \rfloor
%\)
%for $j=0,...,\bgstar N -1.$

\subsubsection{Parameters}

We define three fixed $x$-positions,  $Z_m \le Z_a \le Z_v.$   
These correspond respectively to three $x$-positions on launched nodes: 
a minimum desired position of initial settled fragments, i.e., a minimum desired $z_s(\node)$ ;
the expected value of $z_s(\node,\launchtime(\node)),$  i.e., the expected value 
of $\kappa (h_v(\launchtime(\node)-) +\frac{1}{N});$
the beginning of transient fragments on a launched  node, i.e., $Z_v = 1-\beta_v \kappa.$

Consider an object in position $Z$ and assume that it has $\bgstar N$ settled fragments.  Under continuous standard repair it will reach the
head of the repair queue after an elapsed time $\Tsysrep (1-Z)$ and the expected number of those settled fragments lost during that time is
given by
\[
%\frac{1-e^{-\lambda \Tsysrep (1 - Z)}}{e^{\frac{1}{\bgstar N}} - 1} \,.\bgstar
\bgstar N (1-e^{-\lambda \Tsysrep (1 - Z)})\,.
\]
Assume continuous standard repair, and consider nodes launched at times $k \Delta_t$ for $k=0,...,K-1.$   The expected number of these nodes surviving 
at time $K \Delta_t$ is given by
\[
\sum_{k=0}^{K-1} e^{-\lambda (K-k)\Delta_t} = \frac{1-e^{-\lambda K \Delta_t}}{e^{\lambda \Delta_t} - 1} = \frac{1-e^{-\lambda K \Delta_t}}{e^{\bgstar N} - 1} 
\] 
To simplify notation we introduce
\(
\xi_N = (e^{\frac{1}{\bgstar N}}-1) \bgstar N  = 1 + \frac{1}{2\bgstar N} + \frac{1}{6(\bgstar N)^2} + \ldots \simeq 1\,.
\)
Let us define 
\[
\gamma(Z)  = \frac{1-e^{-\lambda \Tsysrep (1 - Z)}}{\xi_N }\,.
\]
and set $\gamma_a = \gamma(Z_a), \gamma_m = \gamma(Z_m), \gamma_v = \gamma(Z_v).$
Assuming $K(Z) := \Tsysrep (1-Z)/\Delta_t$ is an integer,  the quantity $\gamma(Z)  \bgstar N$ is the expected number of survivors among launches in a time period of length $(1-Z)\Tsysrep.$
By our definitions this implies
\[
Z_a = \frac{\kappa}{N} ( \gamma(Z_v)\bgstar + \frac{1}{N})\,,
\]
and the desired condition $Z_a < Z_v$ reduces to the condition $ \frac{\kappa}{N} ( \gamma(Z_v)\bgstar + \frac{1}{N}) \le Z_v.$
We note the relation
\[
{K(Z)} = -\bgstar N \ln (1-\xi_N  \gamma(Z))\,.
\]
With appropriate parameter choices we will have $\beta > \gamma _m > \gamma_a  > \gamma_v > \beta/2.$

%We will define the various quantities so as to satisfy
%\begin{align}
%\label{eqn:hzmin}
%1-e^{-\lambda \Tsysrep (1 - Z_m)} &= \xi_N \gamma_m \bgstar \betastar
%\\
%\label{eqn:hza}
%1-e^{-\lambda \Tsysrep (1 - Z_{a})} &= \xi_N \gamma_a \bgstar\betastar 
%\\
%\label{eqn:hzv}
%1-e^{-\lambda \Tsysrep (1 - Z_{v})} &= \xi_N \gamma_v \bgstar \betastar =: h^*_v
%\end{align}
%
%Under continuous repair, each of these positions gives rise to an expected node height upon repair, i.e., in $x$-position $1.$
%With this understanding, we will have the following relations,
%\begin{align}
%\label{eqn:Zx}
%\frac{1-e^{-\lambda \Tsysrep (1 - Z_{x})}}{e^{\frac{1}{\bgstar N}} - 1}  = \gamma_x  \betastar  \bgstar N
%\end{align}
%where $x \in \{ m,a,v \}$ 

\subsection{Stopping Time}

%We introduce a stopping time $I_s$ as
%\[
%I_s = \argmin_i \{ f(\tfail_i)N - g_v(\tfail_i)N > N-k-G, \, g_v(\tfail_i)N \not\in  [1,\GAP N] \}
%\]
%
%\begin{lemma}
%We claim that no data loss can occur prior to $\tfail_{I_s},$ i.e.,
%\[
%I_s \le I_f
%\] 
%\end{lemma}
%\begin{IEEEproof}
%If $g_v(\tfail_{I_f})) \not\in  [1,\GAP N]$ then we have $I_s \le I_f$ immediately. 
%Assume now that $g_v(\tfail_{I_f}) N \in  [1,\GAP N].$
%Since  $f(\tfail_{I_f})N > N-k$  we now have  $(f(\tfail_{I_f}) - g_v(\tfail_{I_f}))N > N-k-g_v(\tfail_{I_f}))N \ge N-k-G$ 
%which implies $I_s \le I_f.$
%\end{IEEEproof}
Assuming continuous standard repair, the transitional node completion times are $k \Tnode,$ $k=0,1,2,...$ and we will use the notation
$\node_k$ to indicate the node launched at time $k\Tnode.$
For each $i = 0,1,...$ let us define $J_i = \lfloor s_i/\Tnode \rfloor$ as the index of the node repair immediately preceeding the failure time $s_i.$
Assume $K_m = K(Z_m)$ is integer valued.
Define $\betastar = \beta - 2\gstar.$
%Let $Z_j  = z_s (\node_j).$
%
%Consider the following three sequences of events
%\begin{align*}
%A_i^g & =  \{  g(\tfail_i- )N  \in  [1, \GAP N] \}
%\\
%A_i^Z & = \{ z_s(\node_{J_i- K_m})  \ge  Z_m\}
%\\
%A_i^S & =\{ \survive (\node_{J_i- K_m},J_i\Delta_t)  \le \betastar N \}
%\end{align*}

Consider the following three sequences of events associated to node launches.
\begin{align*}
A_k^g & =  \{  g(k\Tnode+ )N  \in  [1, \GAP N-1] \}
\\
A_k^Z & = \{ z_s(\node_{k- K_m})  \ge  Z_m\}
\\
A_k^S & =\{ \survive (\node_{k- K_m},k\Delta_t)  \le \betastar N \}
\end{align*}
and define the stopping time
\[
I_s = \argmin_i \overline{\{  A_i^g \cup A_i^Z \cup A_i^S\}}\,
\]
as the first launch instance at which at least one of these conditions fails to hold.

\begin{lemma}
No data loss can occur prior to $\tfail_{I_s}.$
\end{lemma}
\begin{IEEEproof}
Let $I_f \Tnode$ denote the last launch time before the first data loss event which occurs at $\tfail_{F}$.  We will show $I_s \le I_f.$
If $g_v(k\Tnode+) \not\in  [1,\GAP N-1]$ or $z_s(\node_{k- K_m}) < Z_m$  for any $k \le I_f$ then we have $I_s \le I_f$ immediately. 
Assume now that $g_v(k\Tnode+)N \in  [1,\GAP N-1]$ and $z_s(\node_{k- K_m}) \ge Z_m$ for all $k \le I_f.$
Since  $f(\tfail_{F} +)N > N-\kcode$  we now have  
\begin{align*}
|\survive(\node_{{I_f} - K_m},{I_f} \Tnode)| 
& \ge
|\survive(\node_{{I_f} - K_m},\tfail_{F}) |
\\
& \ge
(f(\tfail_{F}) - g_v(\tfail_{F}))N 
\\
& >  N-\kcode-g_v(\tfail_{F})N 
\\
&\ge N-\kcode-\GAP = \betastar
\end{align*}
which implies $I_s \le I_f.$
\end{IEEEproof}

%OLD ONE
%\begin{IEEEproof}
%Let $I_f$ denote the last launch time before the first data loss event.  We will show $I_s \le I_f.$
%If $g_v(\tfail_{i}-) \not\in  [1,\GAP N]$ or $Z_{J_i- K_m} < Z_m$  for any $i \le I_f$ then we have $I_s \le I_f$ immediately. 
%Assume now that $g_v(\tfail_{i}) N \in  [1,\GAP N]$ and $Z_{J_i- K_m}\ge Z_m$ for all $i \le I_f.$
%Since  $f(\tfail_{I_f} +)N > N-\kcode$  we now have  
%\begin{align*}
%|\survive(\node_{J_{I_f} - K_m},J(\tfail_{I_f})\Tnode)| 
%& \ge
%|\survive(\node_{J_{I_f} - K_m},\tfail_{I_f}) |
%\\
%& \ge
%(f(\tfail_{I_f}) - g_v(\tfail_{I_f}))N 
%\\
%& >  N-\kcode-g_v(\tfail_{I_f})N 
%\\
%&\ge N-\kcode-\GAP = \betastar
%\end{align*}
%which implies $I_s \le I_f.$
%\end{IEEEproof}

Finally, we show that the expectation of the stopping time $I_s$ is exponentially large in $N.$
\begin{proposition}\label{mainprop}
Assume the stated inital condition with $g_v(t) = \gstar$ and expected transient fragments.
Assume $\beta \le \frac{1}{3},$  $\gstar \le \frac{1}{6} \beta$ and $\gstar N \ge 20.$
Set
\(
\gamma_m = \bgstar \betastar,
\gamma_a = \bgstar^2 \betastar,
\)
and
\(
\gamma_v = \bgstar^3 \betastar
\)
%\begin{align*}
%\gamma_m &= \bgstar \betastar
%\\
%\gamma_a &= \bgstar^2 \betastar
%\\
%\gamma_v &= \bgstar^3 \betastar
%\end{align*}

Then we have
\[
\expectation (I_s) \ge  \frac{1}{8} e^{\frac{3}{8}{\gstar^2}N}
\]
\begin{IEEEproof}
In the appendix we prove the following:
\begin{align}
p(\overline{A_k^g}) &\le 2e^{-\frac{3}{8} \gstar^2 N} \label{akg} \\
p(\overline{A_k^Z}) &\le e^{-\frac{3}{8} \gstar^2 N} \label{akZ} \\
p(\overline{A_k^S}) &\le e^{-\frac{3}{8} \gstar^2 N} \label{akS} \,.
\end{align}
for all $i \le I_s.$
From this we have $p(\overline{\{  A_i^g \cup A_i^Z \cup A_i^S\}}) \le 4  e^{-\frac{3}{8} \gstar^2 N}$
and  an elementary argument now yields the stated result.
\end{IEEEproof}
\end{proposition}
The assumptions on $\beta, \gstar$ and $\gstar N$ are made largely to simplify constants in the proofs.  
They can be relaxed to obtain more general results of the same form.

We first note that setting the three $\gamma$ values determines $Z$ and $K$ and $\Tsysrep.$
In particular we have $K_v = - N \bgstar \ln (1-\xi_N \bgstar^3 \betastar)$ (and which we assume to be integer valued),
which implies $\beta_v =  -  \bgstar \ln (1-\xi_N \bgstar^3 \betastar).$  Letting $N \rightarrow \infty$ we can have $\gstar \rightarrow 0$ and obtain
the asymptotic value $\beta_v = -\ln (1-\beta).$

The value of $\gamma_v$ determines $Z_a$ through $Z_a = \kappa (\gamma_v+\frac{1}{N}).$  The relation between $\gamma_a$ and $Z_a$ then
yields
\[
\lambda \Tsysrep = \bgstar^2 \betastar + \frac{1}{N\bgstar} - \ln (1-\xi_N \bgstar^2 \betastar)
\]
which is asymptotic to $\beta - \ln(1-\beta).$

\subsection{Immediate Repair and Ancillary Repair}

While the above described system achieves arbitrarily large MTDL with what we conjecture are asymptotically optimal repair read rates,
there are various practical drawbacks.
In particular, it may be undesirable to maintain a queue of incomplete nodes with a constant repair rate when an acceleration of the repair process could quickly clear the queue.
Examination of the exponents in the above arguments indicates that quite large systems might be required to enable the described mode of operation with sufficient data protection and accelerated repair could allow smaller systems.
As will be discussed below, the design is also somewhat vulnerable to the probabilistic node failure assumptions.  
In particular, the system depends on the failure of relatively young unsettled nodes to ensure protection from data loss.

In this section we discuss a more practical mode of operation in which we view the storage of transient fragments as
largely opportunistic, intended not to interfere with the ongoing basic liquid-like repair.
In practice the node failure rate is not precisely known and the assumption of exponentially distributed node lifetimes will not hold precisely.
While liquid storage admits delayed repair, it is likely the case that practical repair operations can proceed relatively quickly once a node is declared permanently failed, 
faster than needed according to the node loss rate.
In such a case, assuming an appropriate choice for $\nvfobj,$ the system may reach $g_v(t) N = 0$ frequently.
When $g_v(t)N$ reaches $0$ the standard repair process will simply stop, and it will restart only after a node failure.
Ancillary repair, however, can continue while $g_v(t)N = 0.$  

Without ancillary repair it is still possible for data loss to occur even if node repair is immediate, i.e. even if the repair rate is arbitrarily high.  
Indeed, assume that the node in position $g_v(t)N = 0$ fails repeatedly, i.e., each node failure occurs in position $0.$
Then, eventually we have $h_v(t) N = 0.$ 
Consider a node $\node$ launched under this condition, it has $z_s(\node,\launchtime(\node)) = \frac{\kappa}{N} \simeq 0.$
Now suppose that subsequent to this node launch only fully settled nodes fail.  Upon each subsequent node launch 
$z_s(\node,t)$ will increase by $\frac{\kappa}{N}$ while the node position of $\node$ will increase by $1.$
Since $\frac{N}{\kappa}-1 > N - \kcode$  data loss is inevitable, we will eventually have $h_v(t)N > N - \kcode.$
More generally, if $h_v(t)N$ becomes quite small then the gap between $z_a(\node,t) - z_s(\node,t)$ becomes large, and this leads to the data loss event outlined above.
In this circumstance ancillary repair could increase $z_s(\node,t)$ while leaving $z_a(\node,t)$ fixed, thereby reducing the gap.

If $(h_v(t) + g_v(t))N$ exceeds $N - \kcode$ then data loss occurs.  This could occur even with $g_v(t)N = 0$ is $\beta_v N$ is sufficiently large and
transient fragment carrying nodes do not fail.  It may well be the case in practice that node failure rates are low while the nodes are relatively new and,
in that case, this possibility would become a significant concern.
Let us therefore consider a design in which we choose $\beta_v N  < N -\kcode.$   This implies that $h_v(t)N < N -\kcode$ for all $t$
so that $(h_v(t) + g_v(t))N > N - \kcode$ can occur only with sufficiently large $g_v(t)N.$
By controlling the rate of standard repair, and leaving some additional margin,  one can control the value of $g_v(t)N$ and, with high probability,
keep it sufficiently small.
To give an indication of how this could be accomplished we note the following result.
In an M/D/1 queue with repair time given by $\gamma/\lambda,$ with $\gamma < 1$ the probability that the queue length exeeds $x$
 during a busy period is upper bounded by $e^{-\nu x},$ where $\nu$ solves $e^\nu = 1 + \frac{\nu}{\gamma}.$
 (A proof may be found in the appendix.)
 For example, it $\gamma \simeq 0.313$ (repair time equal to $1/3$ of node failure interarrival times) then $\nu = 2.$
 Hence the probability of exceeding $\delta N$ is less than $e^{-2 \gstar N}.$
 We note the significant improvement of the exponent as compared to the previous section.
 There the critical exponents were of the form $\gstar^2 N.$
 
\begin{comment}
EXAMPLE SYSTEM REMOVED FOR ARXIV UPLOAD
 Let us choose $(\beta_v + \gstar) N = N- \kcode.$
 Let us consider a standard repair policy that repairs nodes in time $R_1/N$ during a busy period as long as $g_v(t)  < \delta_1 < \delta$ but
 if $g_v(t)N$ exceeds $\delta_1 N$ then the repair time is decreased to a time $R_2/N$ and maintained at that time until the busy period ends.
 
 \begin{lemma}
 The probability that $g_v(t)N$ exceeds $\delta N$ during a busy period is less than $e^{-\nu_2}.$
 \end{lemma}
 \begin{IEEEproof}
 \end{IEEEproof}
 
 With appropriate choice of $R_1,R_2,$ and $\delta_1$ we see that the transient fragments will be with high probability confined to a the region with node positions below $N-\kcode.$
 It then follow that data loss can occur only if $z_a(\node,t) - z_s(\node,t)$ was large for for the node in node position $N-\kcode$ at the moment of data loss.
 The gap $z_a(\node,t) - z_s(\node,t)$ can generally be reduced by ancillary repair.
 \end{comment}

\subsection{Use of Repair Regulator for Ancillary Repair}

  In \cite{Luby16} a repair rate regulator was given for the basic liquid storage system for which it was possible to compute strong bounds on the probability of data loss.
 It's basic principle was that objects in the repair queue request repair rates so as to ensure sufficiently small probability of data loss for that object prior to its next repair.
 The scheme is easily adapted to varying queue length since the key parameter is time to repair, as related to node failure rate (which may be estimated).
 That regulator could easily be applied to the settled portion of the queue in the scheme discussed in this section.
 A small modification in the analysis would be required in that objects would be considered `safe' (no repair rate request) if the associated value of
 $f(x,t)$ was in the transient reserve region.  An object in such a position would be protected by the process that ensures the bound on 
 $h_v(t)N + g_v(t)N$ and so would not itself need to request a positive repair rate.

\section{A Complete Virtualization Approach}

We now present an alternative approach in which the entire incomplete portion of the of repair queue is virtualized.
The height of the virtualized queue will be $r = \beta_v N.$ 
Instead of temporarily using certain nodes for saving the overhead fragments, all actual nodes are used simultaneously for both access and overhead.
The amount of overhead needed is $\beta N \simeq \beta_v N/2.$
and the number of actual storage nodes is $N = k_c+ \delta.$
Here $\delta$ represents a margin which protects the system against data loss.

\begin{comment}
\begin{figure}
 \btikzpicture
 \draw[domain=0:10.0,variable=\x,samples=1000] plot ({\x},{0.2*floor(0.5*\x)});
   \fill[red, domain=0:10, variable=\x,samples=1000]   (0,0) -- plot ({\x},{0.1*floor(1*\x)})
   --(10,0) --cycle;
   \fill[blue, domain=0:10, variable=\x,samples=1000]   (0,1.1) -- plot ({\x},{0.1*floor(1*\x)}) -- (10,1.1) -- cycle;
\draw[fill=green] (0,1.1) rectangle (10,5);
\etikzpicture
\end{figure}
\end{comment}

In the complete virtualization approach the objects in the repair queue are maintained in a fixed cyclic order and repaired according to that order in standard repair.
The system requires, however, an ancillary repair process that operates with a different object order.
From the perspective of the repair queue we will view the ancillary repair of objects as happening `in place', meaning their position in the queue is not altered.
Furthermore, we do not separate the two repair processes in time but rather assume a certain amount of synchronization between them.
Unlike the partially virtualized repair method, the method outlined here possesses a unique 'complete' state to which the system will periodically return.
This favorsanalysis of the mode of operation in which the repair rate is generally higher than needed and repair suspends when the complete state is reached.

The objects are partitioned into $N$ groups.  Group membership is determined by object position in the (cyclic) repair queue modulo $N.$
It is convenient, therefore, to assume that $\nobject$ is a multiple of $N,$ so that all groups have precisely the same size and so that
the group definition is invariant under cyclic shift of the repair queue.  This assumption is not critical, but since it simplifies the description we will adopt it.

Each group of objects is uniquely associated to one of the $N$ nodes such that all virtual fragments belonging to objects in the group are stored as transient fragments on the associated node.
When a node fails it is replaced by a new empty node in node position $0$ and the group association of the failed node is transferred to the new node.

When a node fails all objects  lose the settled fragment that had been stored on that node.
In addition, the objects belonging to the group associated to that node each loose all of their transient/virtual fragments.
The intact settled fragment ordering in the virtual repair queue is thereby violated since every $N$th object in the repair queue lost all of its virtual fragments.
Whereas in the partially virtualized method the corresponding objects were advanced in the repair queue, 
in the completely virtualized approach we instead adopt an ancillary repair process that regenerates those missing fragments directly by repairing the objects in the affected group.
The objects otherwise maintain their place in the queue and for each object in the group the lost transient fragments are simply regenerated and stored on the replacement node.
In general this involves regenerating for each object in the group one settled fragment and a varying number of transient fragments.
On average only $\beta_v N/2$ fragments are repaired per object, so the repair efficiency of this ancillary process is less than that of the main repair process by a factor of two.
If the overhead $\beta$ is small, though, then this represents a small portion ($O(\beta)$) of the total needed repair.

\subsection{The Complete State}

It is most convenient to describe the system by first describing the complete state (in which repair suspends).
In the complete state each object has $N$ settled access fragments stored one each on the $N$ access nodes.
The virtualized portion of the repair queue has a staircase form, i.e., an asymptotically linear boundary.
In particular we have $f^*(x)N =   \lfloor -(\beta_v N) (1- x) \rfloor,$
where the superscript $^*$ indicates the complete state and we use the same definition of $f$ as in the previous section.
This function is a step function and each step has a width of $\frac{1}{\beta_v N} \nobject$ objects.
See Fig.  \ref{fig:complete} for an example.
Note that the $x$-axis is now extended beyond $x=1$ with $x > 1$ representing the overhead portion of the storage capacity.

It is convenient (but not critical) to assume that $\frac{\nobject}{\beta_v N^2}$ is an integer. In that case each group of objects comprises exactly the same number of virtual fragments.
Since the incomplete repair queue is entirely virtualized, the actual overhead of the system is not $\beta_v$ but, approximately, $\beta_v /2.$
A careful check of the definition of $f^*$ shows that the virtualized portion of the repair queue actually includes one completed node.
It is possible in the complete virtualization approach to include zero or more than one completed nodes in the virtualization, e.g. by  setting
 $f^*(x)N =   \lfloor -s (1- x) -  (\beta_v N -s) \rfloor$ for some $s < \beta_v N,$ and adding more complete virtual nodes would provide an additional buffer against bursty node losses at the cost of additional overhead, but, to simplify the
presentation, we will not develop these variations.

\subsubsection{Node Failure in the Complete State}
Consider a node failure while the system is in the complete state.
One group in the virtualized portion of the repair group is erased.
A new empty node is added to the system in node position $0$ in the repair queue.
The surviving nodes in positions below the failed node are all advanced by $1$ in their node positions.

Conceptually, all virtual fragments are incremented by $1$ also in the node ordering.
Note that this means that the top virtual node now coincides with the new empty physical node, and the associated transient fragments will be copied to the new physical node.
An ancillary repair job is simultaneously commenced to regenerate the erased transient fragments, all of which will be written to the new node.
In addition, each ancillary object repair regenerates the one missing settled fragment associated to the lost node.
Because we assume one complete virtual node in the complete state, the new physical node will be complete once the ancillary repair job and the transient copying are complete.
This does not, however, in itself recreate the complete state.
In order to reach the complete state the standard repair process, with the virtual/transient fragments being written to their associated nodes,
must also complete $\frac{\nobject}{\beta_v N}$ standard object repairs.
Note that a small number of objects are scheduled for repair in both the ancillary and the regular repair process.
Clearly, only one repair for those objects is required.

%Assuming distributed repair computation, that node would be the best location to perform the repair computation.

\subsubsection{Further Node Failure}

If additional nodes fail before reaching the complete state, then, for each additional failure, another $\frac{\nobject}{\beta_v N}$ standard object repairs are scheduled
along with the ancillary repair for each failed node.   Thus, each repair job involves $\frac{\nobject}{\beta_v N}$ standard object repairsat most and fewer than
$\frac{\nobject}{N}$ ancillary object repairs.
Each node repair also entails the copying of transient fragments to the transitional node.
When the transitional node is complete the storage used for those transient fragments can be released.
We will consider two possible relationships between standard repair and ancillary repair.  
The first and simplest is to consider both repairs as associated to the transitional node repair. 
 In this approach both types of repair are tied together in a single repair function and tied to the repair of the transitional node.
 When the transitional node completes its standard repair it may not be fully settled since the objects belonging to groups associated to other
failed nodes waiting to advance to the transitional node for repair will be missing fragments associated to those nodes.
In the second approach the ancillary repairs are given priority.  Since ancillary repairs per node are smaller (order $\beta$) than the standard repairs
and this improves the resiliency of the system for a given $\delta,$ this is likely the more practial approach.
It does, however, lead to a more complicated analysis of the performance of the system.

\subsubsection{Atomic Nature of Transitional Node Repair}
 The transient fragments that are copied to the transitional node remain intact until the transitional node is complete.
The number of settled fragments regenerated by the ancillary repair process for the transitional node is less than $\frac{\nobject}{N}.$
With a negligible fraction of the storage capacity (roughly $N^{-2}$) these fragments could be written to both the transitional node and temporarily copied to other nodes.
If the transitional node then fails while it is being written, it could be reconstituted with copying alone.
In part to simplify the analysis, we will assume that transitional nodes cannot fail. 
Using the above mechanism, this could be effectively realized in an actual system with a small amount of additional overhead.

\begin{figure}
 \begin{tikzpicture}
 \draw[domain=0:10.0,variable=\x,samples=1000] plot ({\x},{0.2*floor(0.5*\x)});
   \fill[red, domain=0:10, variable=\x,samples=1000]   (0,0) -- plot ({\x},{0.1*floor(1*\x)})   --(10,0) --cycle;
   \fill[cyan, domain=0:10, variable=\x,samples=1000]   (0,1.0) -- plot ({\x},{0.1*floor(1*\x)}) -- (10,1.0) -- cycle;
\draw[fill=green] (0,1.0) rectangle (10,5);

%indicate objects in group black
\foreach \z in {0,...,19}
\draw[black!40!green,fill=black!40!green,samples=5000] (0.2+0.5*\z,0.1) rectangle ++(0.0000125,5-0.1); %should be 0.0125 wide but is uneven

%transient fragments
\draw[fill=cyan] (10,5) rectangle ++(1.25,-4);

%rectangle of indicated node
\draw[black!60!green,line width=0.01mm,samples=1000] (0,4) rectangle ++(11.25,-0.1);

%virtual bounding box
\draw[samples=1000] (0,0) rectangle ++(10.0,1.0);

%indicated virtual fragments
\foreach \z in {0,...,19}
\draw[blue,fill=blue] (0.2+0.5*\z,0.01) rectangle ++(0.0000125,1); %should be 0.0125 wide

\draw[blue,fill=blue] (10,4) rectangle ++(1.25,-0.1);
   \fill[red, domain=0:10, variable=\x,samples=1000]   (0,0) -- plot ({\x},{0.1*floor(1*\x)})   --(10,0) --cycle;
   
   \node[below] (v2) at (-0.5,0) {};
\node[above] (v3) at (-0.5,1) {};
\node[below] (v3b) at (-0.5,1) {};
\node[above] (v4) at (-0.5,5) {};
\draw[thick,<->]  (v2) -- (v3) node[midway,left] { Virtual Nodes };
\draw[thick,<->]  (v3b) -- (v4) node[midway,left] { Access Nodes };
\end{tikzpicture}
\caption{\label{fig:complete} Repair Queue in Complete State.  We assume 40 access nodes with $\beta_v N = 10.$
The area associated to one node is indicated, along with its associated group. }
\end{figure}
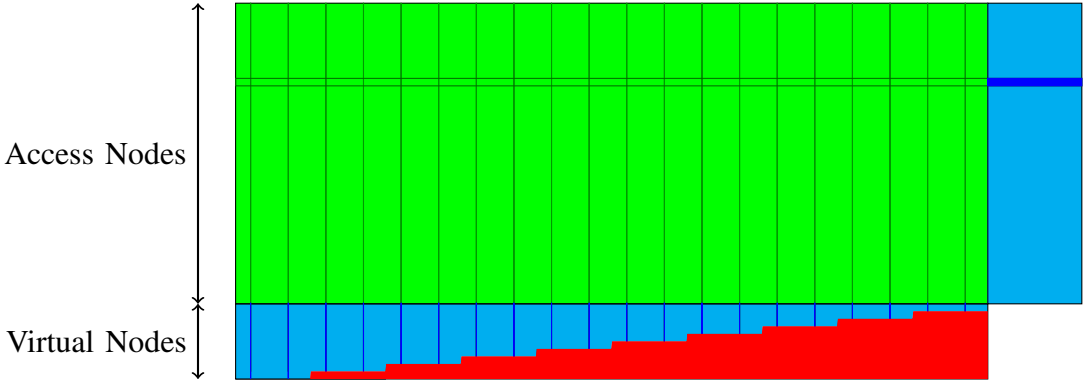

\begin{figure}
 \begin{tikzpicture}
%10 10 nodes    
% x is from 0 to 1
%bv N = 10
 \draw[domain=0:10.0,variable=\x,samples=1000] plot ({\x},{0.2*floor(0.5*\x)});
   \fill[red, domain=0:10, variable=\x,samples=1000]   (0,0) -- plot ({\x},{0.1*floor(1*\x)})   --(10,0) --cycle;
   \fill[cyan, domain=0:10, variable=\x,samples=1000]   (0,1.1) -- plot ({\x},{0.1+0.1*floor(1*\x)}) -- (10,1.1) -- cycle;
\draw[fill=green] (0,1.1) rectangle (10,5);

%indicate objects in group black
\foreach \z in {0,...,19}
\draw[black!40!green,fill=black!40!green,samples=5000] (0.2+0.5*\z,0.1) rectangle ++(0.0000125,5-0.1); %should be 0.0125 wide
%transient nodes
\draw[fill=cyan] (10,5) rectangle ++(1.25,-3.9);

%bounding box of top virtual
\draw (0,1.1) rectangle ++(11.25,-0.1);

%rectangle of transient nodes to be copied
\draw[blue,fill=blue] (10,5) rectangle ++(0.25,-3.9);
%rectangle of vitrual nodes to be copied
\draw[blue,fill=blue] (0,1.1) rectangle ++(10,-0.1);

%white erased virtuals
\foreach \z in {0,...,19}
\draw[white,fill=white] (0.2+0.5*\z,0) rectangle ++(0.0000125,1.1);
%\draw[blue,fill=blue] (10,4) rectangle ++(1.25,-0.1);

%redraw red
 \fill[red, domain=0:10, variable=\x,samples=1000]   (0,0) -- plot ({\x},{0.1+0.1*floor(1*\x)})   --(10,0) --cycle;
 
%virtual bounding box
\draw[samples=1000] (0,0) rectangle ++(10.0,1.0);

%labels  
\node[below] (v2) at (-0.5,0) {};
\node[above] (v3) at (-0.5,1) {};
\node[below] (v3b) at (-0.5,1) {};
\node[above] (v4) at (-0.5,5) {};
\draw[thick,<->]  (v2) -- (v3) node[midway,left] { Virtual Nodes };
\draw[thick,<->]  (v3b) -- (v4) node[midway,left] { Access Nodes };
\end{tikzpicture}
\caption{Result of  Failure of Indicated node from Fig. \ref{fig:complete}. Transient fragments available for copying are indicated in dark blue.
The white strips indicate erased fragments.}
\end{figure}
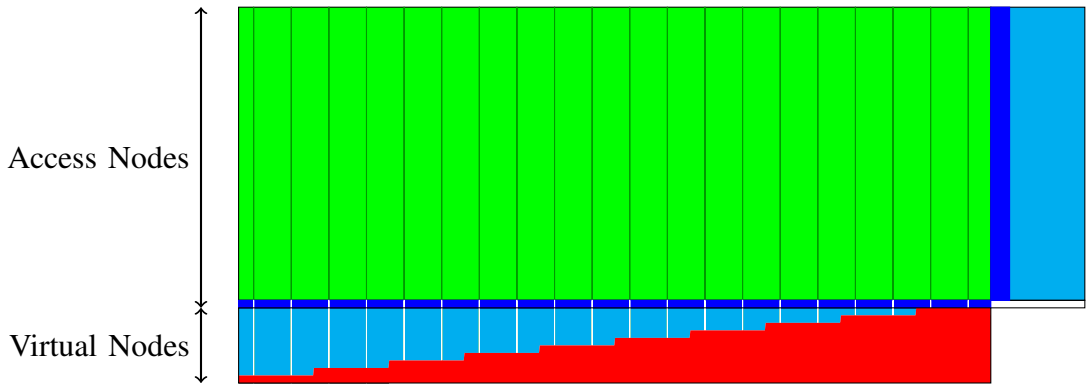

\begin{figure}
 \begin{tikzpicture}
%10 10 nodes    
% x is from 0 to 1
%bv N = 10
 \draw[domain=0:10.0,variable=\x,samples=1000] plot ({\x},{0.2*floor(0.5*\x)});
  \fill[red, domain=0:10, variable=\x,samples=1000]   (0,0) -- plot ({\x},{0.3+0.1*floor(1*(\x-0.7))})   --(10,0) --cycle;
  \fill[cyan, domain=0:10, variable=\x,samples=1000]   (0,1.3) -- plot ({\x},{0.3+0.1*floor(1*(\x-0.7))}) -- (10,1.3) -- cycle;
\draw[fill=green] (0,1.3) rectangle (10,5);

%indicate objects in group black
\foreach \z in {0,...,19}
\draw[black!40!green,fill=black!40!green,samples=5000] (0.4+0.5*\z,0.1) rectangle ++(0.0000125,5-0.1); %should be 0.0125 wide
%transient nodes
\draw[fill=cyan] (10,5) rectangle ++(1.25,-3.7);

%rectangle of transient nodes to be copied
\draw[blue,fill=blue] (10,5) rectangle ++(0.25,-3.7);

%rectangle of vitrual nodes to be copied
\draw[blue,fill=blue] (0,1.2) rectangle ++(10,0.1);
\draw[blue] (0,1.1) rectangle ++(10,0.1);
\draw[blue] (0,1.0) rectangle ++(10,0.1);

%white erased virtuals
\foreach \z in {0,...,19}
\draw[white,fill=white] (0.4+0.5*\z,0.2) rectangle ++(0.0000125,1.1);
\foreach \z in {0,...,13}
\draw[black!40!green,fill=black!40!green] (0.4+0.5*\z,0.2) rectangle ++(0.0000125,1.1);
\foreach \z in {0,...,19}
\draw[yellow,fill=orange] (0.1+0.5*\z,0.2) rectangle ++(0.0000125,1.1);
\foreach \z in {0,...,19}
\draw[white,fill=white] (0.25+0.5*\z,0.2) rectangle ++(0.0000125,1.1);

%repaired transients
\draw[black!40!green,fill=black!40!green] (10.0,1.2) rectangle ++(1.05,0.1);
%\draw[blue,fill=blue] (10,4) rectangle ++(1.25,-0.1);

%redraw red and virtual-actuals
 \fill[red, domain=0:10, variable=\x,samples=1000]   (0,0) -- plot ({\x},{0.3+0.1*floor(1*(\x-0.7))})   --(10,0) --cycle;

%repaired fragments
%\draw[fill=green] (0.0,1.2) rectangle ++(0.7,0.1);
%\draw[fill=green] (0.0,1.1) rectangle ++(0.4,0.1);

%rectangle of new transient nodes to be copied (40%)
\draw[cyan,fill=cyan] (11.25,5) rectangle ++(0.125,-3.9);
\draw[cyan,fill=cyan] (11.375,5-0.1) rectangle ++(0.125,-0.2);
\draw[cyan,fill=cyan] (11.375,5-0.35) rectangle ++(0.125,-0.25);
\draw[cyan,fill=cyan] (11.375,5-0.9) rectangle ++(0.125,-0.25);
\draw[cyan,fill=cyan] (11.375,5-1.2) rectangle ++(0.125,-0.3);
\draw[cyan,fill=cyan] (11.375,5-1.6) rectangle ++(0.125,-0.1);
\draw[cyan,fill=cyan] (11.375,5-2.43) rectangle ++(0.125,-0.1);
\draw[cyan,fill=cyan] (11.375,5-3.2) rectangle ++(0.125,-0.4);

%virtual bounding box
\draw[black] (0,0) rectangle ++(10.0,1.0);
\draw[black] (11.25,5) rectangle ++(0.25,-4.0);

%bounding box of top virtual
\draw (0,1.3) rectangle ++(11.25,-0.1);
\draw (0,1.2) rectangle ++(11.25,-0.1);
\draw (0,1.1) rectangle ++(11.25,-0.1);

%labels  
\node[below] (v2) at (-0.5,0.0) {};
\node[above] (v3) at (-0.5,1) {};
\node[below] (v3b) at (-0.5,1.35) {};
\node[above] (v4) at (-0.5,5) {};
\draw[thick,<->]  (v2) -- (v3) node[midway,left] { Virtual Nodes };
\draw[thick,<->]  (v3b) -- (v4) node[midway,left] { Access Nodes };
\end{tikzpicture}
\caption{Result of Three Node Failures.  The repair of the first failure is 70\% complete. 
After 25\% of the repair was complete a second node failed and after 60\% a third node failed.
Dark green indicates repaired fragments.
Here we show the delayed copy model where transient fragments for transitional and virtual nodes are lost when another node fails.
}
\end{figure}
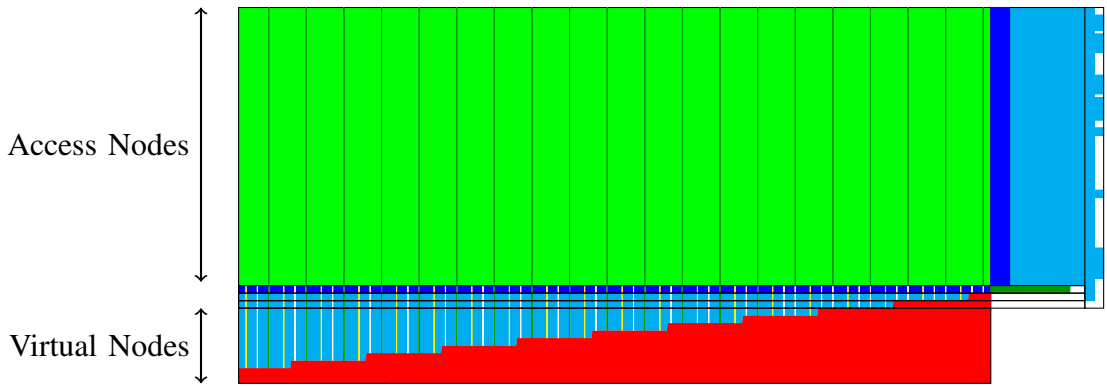

%\begin{figure}
%\includegraphics{FigLubyMove.png}
%%\input{FigLubyMoveTikz.tex}
%\caption{Result of Node Failure and Data Movement}
%\end{figure}

\subsection{Analysis}

We define $g_v(t)$ as in the partially virtualized case, i.e., $g_v(t)N$ denotes the node position immediately above the transitional node, assuming $g_v(t)N > 0.$
Here $g_v(t) = 0$ implies the complete state with repair suspended.
Let us first consider the case where ancillary repair is synchronized with standard repair.
With the assumption of no transitional node failure, the system behaves as single server queue with $g_v(t)$ representing the number
of users in the queuing system at time $t.$
Under fixed repair time this is precisely a $M/D/1//N$ machine interference problem.
In particular, the arrival rate decreases as $g_v(t)$ increases.  In practice, however, the value of $g_v(t)$ would be kept small,
and in fact cannot exceed $\delta$ without data loss.
The repair rate of the system could be adjusted to increase as $g_v(t)$ increases to control the probability of data loss.
%Any decrease above the nominal repair time $1/\lambda$ gives an exponential decrease in the tail probabilities of the queue.

The number of erased access fragments for objects in the system can depend on the group of that object.
All objects are missing the $g_v(t)N$ fragments associated to the standard repair queue.
In addition, objects belonging to groups associated to failed nodes awaiting repair will be missing additional fragments.
When a node fails all of the virtual fragments associated to that node's group are lost.  
They will not be recovered by the ancillary repair until that node becomes and completes as the transitional node.
All nodes that were in the repair queue at the time of failure will be missing those fragments until the node completes as a transitional node.
If an object belongs to a group associated to a failed node that is in the node repair queue, then it is missing fragments for all nodes that were ine
node repair at the time of its failure. 
Hence the number of erased fragments for an object is $g_v(t) N + E$ where $E$ is $0$ is the object's group-associated node in not in the node repair queue 
and is otherwise equal to the number of nodes that were in the node repair queue at the time of the node's failure.
It follows that as long as $g_v(t)N  < \frac{1}{2}\delta N$ is maintained then no data loss does occurs. 
Thus, controlling the repair rate ensure $g_v(t) < \frac{1}{2}\delta$ provides data integrity.

For and M/D/1 queue with arrival rate $\lambda'$ and service time $D' = \gamma/\lambda'$ with $\gamma < 1,$ we show in Appendix \ref{appMD1}
that the probability of the queue exceeding $m$  during a busy period is less than $e^{-\nu m}.$
Excursion probabilities are only smaller in the finite population case, (where the arrival rate decreases with queue size).
The probability during a busy period that $g_v(t)  > \frac{1}{2} \delta N - 1$ is less that $e^{-\nu \frac{1}{2}\gstar N }$
where $\nu$ is given by $e^{\nu} = 1 + \frac{\nu}{\gamma}$ where the time for a node repair is $\gamma /(\lambda N).$
For example, if $\gamma \simeq 0.31$ then $\nu = 2.$
Thus the expected number of busy periods until data loss is at least $e^{\nu (\gstar N - 1)}-1.$

The expected length of each busy period until this occurs as at least $\frac{1}{1-\mu}$ where $\mu = 1/(\bgstar N \lambda D) = 1/(\bgstar \gamma).$
Thus, MTTDL is greater than $(e^{\nu \gstar N}-1) (\frac{1}{\bgstar \gamma} D + \frac{1}{\lambda N}).$
It follows that we have MTTDL $\rightarrow \infty$ for any fixed $\gamma < 1.$  Moreover, we can have $\gamma \rightarrow 1$ as $N \rightarrow \infty.$
In this asymptote the read repair rate is given by $(\lambda \Tsysrep)^{-1} (2\beta + O(\beta^2))$ which is optimal to first order in $\beta.$

In the case where ancillary repair is given priority over standard repair the repair process can be viewed as a two stage queuing system where the
first stage performs the ancillary repair and the second stage performs the standard repair.   We may assume a fixed overall repair rate that operates
on one stage at a time, or a variable rate system that accelerates repair as queue size increases.

\appendices

%\input{AlgorithmWtestappendix.tex}


\section{Proof of Proposition \ref{mainprop}}

To facilitate the proof we define an non-terminating version of the repair process.
We assume a system that has $M = \bgstar N$ nodes at time $t=0,$
according to  the stated initial condition. 
At each time $k\Tnode,$ $k=1,2,...$ a node is launched into, i.e. added to, the system.
The duration from launch to failure of a node is an independent exponentially distributed random variable with rate $\lambda.$
We will use the notation $\node_k$ to indicate the node launched at time $k\Tnode.$

To model the node failure process in an alternate way we adopt a point Poisson process with rate $\lambda N.$
The arrival times of the Poisson process will be denoted $\tfail_i,$ $i=0,1,2,...$
Here, $\tfail_0 = 0$ and for $i>0$ the differences $\tfail_i - \tfail_{i-1}$ are i.i.d. exponential random variables with rate $\lambda N.$
For each $i > 0$ we further adopt a uniformly random independent random variable $Y_i \in [0:N-1].$
We interpret this to mean that the node in node position $g_v(\tfail_i )N+Y_i$ is affected by the failure event associated to the Poisson process.   
If $Y_i + g_v(\tfail_i  )N \ge N$ then no node failure actually occurs and the system is unaffected, otherwise
the node in node-position $g_v(\tfail_i )N+Y_i$ fails at time $\tfail_i.$

%Node failure is derived from a Poisson process and the time to failure for any active node is an exponentially distributed random variable with rate $\lambda.$
 
 \subsubsection{Bounds on $g_v$)}
Let $M_k$ denote the number of surviving launched nodes in the system at time $k\Tnode +,$ i.e., immediately after the $k$th node launch.
Let us introduce the notation $q = (1 - e^{-\lambda \Tnode})$ and $\bar{q} = 1 -q.$
A node surviving at time $k\Tnode$ will fail by time $(k+1)\Tnode$ with probability $q.$
We note that
\[
q^{-1} =( 1 - e^{-\frac{1}{\bgstar N}})^{-1} = \bgstar N + \epsilon_N
\]
where $\epsilon_N \in (0,1)$ for $\bgstar N \ge 1.$

\begin{lemma}[Proof of \eqref{akg}]
For $k \ge 1$ we have 
\begin{align}
\prob ( M_k > N-1) &\le  e^{-\frac{3\gstar^2}{8}N} \label{upperM}
\\
\prob (M_k < (1-2\gstar)N ) & \le e^{-\frac{3\gstar^2}{8} N} \label{lowerM}
\end{align}
\end{lemma}
\begin{IEEEproof}
For $k \ge 1$ the number of surviving launched nodes at time $k\Tnode$ can be written as a sum of independent Bernoulli random variables.
\[
M_k = \sum_{j=1}^{\bgstar N} \tilde{b}_j +  \sum_{j=0}^{k-1}b_j
\]
where $\tilde{b}_j$ indicates the survival of the $j$th initial node
and $b_j$ indicates the survival of the node launched at time $(k-j) \Tnode.$
It follows that 
\(
\tilde{p}_j := \expectation (\tilde{b}_j) = \bar{q}^k
\)
and
\(
p_j := \expectation ({b}_j) = \bar{q}^{j}
\)
and we obtain
\begin{align*}
\expectation(M_k) &= \bgstar N \bar{q}^k + \sum_{j=0}^{k-1} \bar{q}^{j}
\\ &
= \bgstar N \bar{q}^k + \frac{1- \bar{q}^{k}}{q}
\\ &
= \bgstar N  + \epsilon_{N,k}
\end{align*}
where $\epsilon_{N,k} = \epsilon_N (1- \bar{q}^{k}) \in (0,1).$

Define $E = \sum_{j=1}^{\bgstar N} \min (\tilde{p}_j,1-\tilde{p}_j) +  \sum_{j=0}^{k-1} \min ({p}_j,1-{p}_j)$ and
set $W = \lceil \frac{-\ln 2}{\ln \bar{q}}\rceil = \lceil (\ln 2) \bgstar N \rceil.$ 
We claim $E \le W.$ Assume first that $k \ge  W.$ Then
\begin{align*}
E
&\le
(\bgstar N)\bar{q}^k + \sum_{j=0}^{W-1} (1 - \bar{q}^j) + \sum_{j=W}^{k-1} \bar{q}^j
\\
&=
(\bgstar N)\bar{q}^k +W - \frac{1 - \bar{q}^W}{q} +   \frac{ \bar{q}^W - \bar{q}^k }{q}
\\
&=
(\bgstar N - \frac{1}{q})(\bar{q}^k) + W- \frac{1 - 2\bar{q}^W}{q}
%&=
%(\bgstar N - \frac{1}{q})(\bar{q}^k) + W- \frac{1 - (1-q)^{\xi}}{q}
\\
&\le
W
\end{align*}
where for the last step we use 
$0 < \bar{q}^W\le  \bar{q}^{\frac{-\ln 2}{\ln\bar{q}}} = \frac{1}{2}.$
The argument for $k \le W$ is similar and for that case we actually obtain the slightly stronger bound  $E \le k \le W.$

% Assume $k \le W.$Then
%\begin{align*}
%&\sum_{j=1}^{\bgstar N} \min (\tilde{p}_j,1-\tilde{p}_j) +  \sum_{j=0}^{k-1} \min ({p}_j,1-{p}_j)
%\\
%&\le
%(\bgstar N)(1-\bar{q}^k) + \sum_{j=0}^{k-1} (1 - \bar{q}^j)
%\\
%&=
%(\bgstar N)(1-\bar{q}^k) +k - \frac{1 - \bar{q}^k}{q}
%\\
%&=
%(\bgstar N - \frac{1}{q})(1-\bar{q}^k) + k
%\\
%&\le
%k
%\end{align*}
%$\bar{q}^W \in [0,\frac{1}{2}]$ as can be seen from
%$\bar{q}^W = e^{-\ln (2) +  \xi \ln \bar{q}} = \frac{1}{2}\bar{q}^{\xi} ,$ where $\xi = W + \frac{\ln 2}{\ln \bar{q}} \in [0,1].$

Applying Lemma \eqref{chernoff} (Chernoff bound) we now obtain
\[
\prob(M_k > N-1) \le e^\frac{-3\eta^2}{8 W}
\]
with $\eta = N-1-(\bgstar N + \epsilon_{N,k}) = \gstar N - 1 - \epsilon_{N,k}.$
Noting that $W \le \bgstar N \ln (2)+1$ and using the conditions of Proposition \ref{mainprop} we have
\begin{align*}
(\frac{\eta}{\gstar N})^2
&
=
(1 - \frac{1+\epsilon_{N,k}}{\gstar N})^2
\\
&
\ge
0.81
\\
&
\ge
\bgstar (\ln 2) + \frac{1}{N}
\\
&\ge
\frac{W}{N}
\end{align*}
and we see that \eqref{upperM} holds.

Similarly, Lemma \eqref{chernoff} gives
\[
 \prob(M_k < (1-2\gstar) N ) \le e^{-\frac{3\eta^2}{8 W}}
\]
with $\eta =\gstar N + \epsilon_{N,k}$ so  \eqref{lowerM} follows from
$W \le \ln(2) \bgstar N +1 \le N.$
\end{IEEEproof}

\subsubsection{Bounds on Settled Nodes}

We now consider the number of nodes lost in a fixed time interval of length $K \Tnode$ from among the launched nodes present at node launch time $(k-K)\Tnode.$
%In particular we are interested in the probability that the number of node failures exceeds $\betastar N.$
The expected value takes the form $M (1-q)$ where $M$ is an initial number of nodes and $q$ is the probability of one such node failing in the given time interval.

\begin{lemma}
Assume $\betastar -  (1-q^K)  > 0$ (with $q = e^{-\lambda  \Tnode}$).  Let $F$ be the number of settled nodes present at time $(k-K)\Tnode$ that
fail by time $k\Tnode.$
\begin{align}
\ln \prob(F \ge \betastar N) &\le
{- \frac{3(\betastar N - N (1- \bar q^K))^2}{8 N (1- \bar q^K)}}
\label{lossbound}
\end{align}
\end{lemma}
\begin{IEEEproof}
Assume first that $k-K \ge 0$ and that there are $M$ launched nodes at time $(k-K)\Tnode.$
The number of these nodes that fail by time $k\Tnode$ is a random variable $F$ given by $F = \sum_{i=1}^M (1 - b_i)$ where the $b_i$ are i.i.d. Bernoulli with
$\expectation (b_i) = \bar q^K$ and so under these conditions $\expectation(F) = M (1- \bar q^K).$
Applying Lemma \ref{chernoff} we obtain
\begin{align}
\ln \prob(F \ge \betastar N) &\le {- \frac{3(\betastar N - M (1- \bar q^K))^2}{ 8M (1- \bar q^K)}}\,.
\end{align}
Under the stated assumptions the quantity on the right is decreasing in $M$ for $M\le N,$
and increasing in $\bar q$ so the desired result holds for this case.

Assume now that $k-K < 0$ we have $F = \sum_{i=1}^M (1 - b_i)$ where $b_i$ are i.i.d. Bernoulli with
$\expectation (b_i) = q^k,$  and so $\expectation(F) = M (1- q^k)$
and $M$ is the number of initial nodes surviving from time $( k-K)\Tnode,$ so $M = \lceil S_{K-k} \rceil  < N.$
It follows that the above bound holds in this case as well.
\end{IEEEproof}

\subsubsection{Bounds on Transients}

We now consider the node survival process over relatively small launch windows.
In particular we consider how many nodes launched from $(k-K)\Tnode$ until $k\Tnode$ survive 
at time $k\Tnode\!+.$  We are generally interested in the case $K = K_v$ and in that case the number of survivors
is equal to $h_v (k\Tnode\!-) N + 1,$ which in turn gives the number of initial settled fragments placed on the transitional node.

\begin{lemma}\label{lem:Kbound}
Assume $K$ satisfies $\bar{q}^K > \frac{1}{2}$ and $K q < 1.$
Then for all $k \ge 0$ we have
\begin{align}
%\prob (\survive_{k-K}^k \ge \expectation(\survive_{k-K}^k) + \eta ) &\le e^{-\frac{3\eta^2}{8\binom{K+1}{2}q}}
%\\
\prob (\survive_{k-K}^k \le \expectation(\survive_{k-K}^k) - \eta ) &\le e^{-\frac{3\eta^2}{8\binom{K+1}{2}q}}
\end{align}
\end{lemma}
\begin{IEEEproof}
If $k - K \ge 0$ then $\survive_{k-K}^k$  is independent of the initial condition and is given by
\(
 \sum_{j=0}^{K} b_j
\)
where $b_j$ indicates the survival of the node launched at time $(k-j) \Tnode.$ 
We have $p_i := \prob (b_j = 1) = \bar{q}^{j}$
so
\(
\expectation ( \sum_{j=0}^{K} b_j) = \sum_{j=0}^{K} p_j = \frac{1 -  \bar{q}^{K+1}}{ q }\,.
\)
With the assumptions on $K$ we have 
\begin{align*}
 \frac{1 -  (1-q)^{K+1}}{ q }\, &=   \frac{{(K+1)}q - \binom{K+1}{2}q^2 + ...}{ q }
 \\
 &\ge
  (K+1) - \binom{K+1}{2}q\,.
\end{align*}
and it now follows that
\(
 \sum_{j=0}^{K} (1- p_{j}) \le  \binom{K+1}{2}q\,.
\)
%\begin{align*}
% \sum_{j=0}^{K} (1- p_{j}) & = 
%  K+1 - \frac{1 -  (1-q)^{K+1}}{ q }
%  \\
% &  = 
%  K+1 - \frac{1 -  (1 - (K+1) q +\binom{K+1}{2}q^2 \ldots)}{ q }
%    \\
% &  = 
%{\binom{K+1}{2}q- \binom{K+1}{3}q^2 \ldots}
%\\
% &  \le 
%\binom{K+1}{2}q
%\end{align*}

In the case  $k-K < 0$
then $\survive_{k-K}^k$ is given by 
\(
\sum_{j=1}^{\lceil S_{K-k} \rceil} \tilde{b}_j + \sum_{j=0}^k b_j
\)
where $\tilde{p}_i = \prob(\tilde{b}_j=1) = \bar{q}^k.$
Now
\begin{align*}
\sum_{j=1}^{\lceil S_{K-k} \rceil} \tilde{p}_j + \sum_{j=0}^k p_j
%& =
%{\lceil S_{K-k} \rceil} \bar{q}^k + \sum_{j=1}^k \bar{q}^{k-j}
%\\
& \ge
{S_{K-k} } \bar{q}^k + \sum_{j=1}^k \bar{q}^{k-j}
%\\
%&=
% \sum_{j=k}^{K} \bar{q}^{j} + \sum_{j=0}^{k-1} \bar{q}^{j}
\\&
=
 \frac{1 -  \bar{q}^{K+1}}{ q }
 \\&
 \ge K+1 - \binom{K+1}{2}q
\end{align*}
and, since $\lceil S_{K-k}\rceil + k < K+1$ we obtain
\(
\sum_{j=1}^{\lceil S_{K-k} \rceil} (1-\tilde{p}_j) + \sum_{j=0}^k (1-p_j)  \le \binom{K+1}{2}q \,.
\)

%\begin{align*}
%\binom{K+1}{2}q - \xi \bar{q}^k - (K+1 - k - \lceil S_{K-k}\rceil)
%\le \binom{K+1}{2}q - \xi \bar{q}^k
%\end{align*}

Applying the Chernoff bounds, Lemma \ref{chernoffsum} we now obtain the desired result.
\end{IEEEproof}

\subsubsection{Application of Bounds}

We note that for $x \in [0,1)$ we have $x \le -\ln (1-x) \le  x(1+\frac{1}{2}\frac{x}{1-x}).$
\begin{lemma}\label{Kbound}
Assume $\bgstar^2 \betastar N \ge 10,$ that $\bgstar \beta N > 20,$ that $\bgstar \gstar N \ge 2.1,$ that $\beta \le \frac{1}{3},$ and that $\gamma(Z) \le \bgstar\betastar.$  Then
\[
\binom{K(Z)+1}{2} \frac{q}{N} \le \gamma^2(Z)
\]
\end{lemma}
\begin{IEEEproof}
To simplify notation we will suppress dependence on $Z.$
First, we note the bound
\[
\binom{K+1}{2} q \le \frac{(K+1.5)^2}{2}\frac{1}{\bgstar N}
\]
Since $\xi_N \gamma \le \xi_N \bgstar\betastar \le \betastar \le \frac{1}{3}$
we have
\[
\frac{K}{N} = -\bgstar \ln (1-\xi_N  \gamma) \le 1.25 \bgstar\xi_N \gamma
\]
which we can combine with $1 \le N \bgstar\xi_N \gamma$ to obtain
(Since $\gamma \ge \beta/2$ and $\beta N \ge 20$ we have $N \bgstar \xi_N \gamma \ge 10$ )
\[
{K+1.5}  \le \sqrt{2}\, \bgstar\xi_N \gamma  N\,.
\]
Combining the above we now have
\[
\binom{K+1}{2} q \le  \bgstar (\xi_N \gamma)^2 N  \le \gamma^2 N
\]
where we used $\bgstar \xi_N^2  \le 1-\gstar + \frac{2}{N} \le 1.$ 
\end{IEEEproof}

\begin{lemma}[Proof of \eqref{akS}]
\[
\prob(S_{k-K_m}^k \ge \betastar M) \le e^{-\frac{3}{8} \gstar^2 N}
\]
\end{lemma}
\begin{IEEEproof}
Now we consider $S_{k-K_m}^k.$
Note that $\expectation |\survive (\node, \launchid(\node)+K\Delta_t)| = \gamma_m N.$
Hence $\betastar N - \expectation |\survive (\node, (\launchid(\node)+K_m)\Delta_t)|  = (\betastar-\gamma_m) N = \gstar \betastar N.$
Thus, applying Lemma \ref{lem:Kbound} we see that the lemma will follows from
\begin{align*}
\frac{( \betastar-\gamma_m)^2 N^2}{\binom{K_m+1}{2} q} 
& =  \frac{\gstar^2 \betastar^2  N^2}{\binom{K_m+1}{2} q} 
\\
& \ge
\frac{\gstar^2 \betastar^2  N^2}{\gamma_m^2} 
\\
&\ge \gstar^2 N 
\end{align*}
%\[
%\frac{( \betastar-\gamma_m)^2 N^2}{\binom{K_m+1}{2} q} \ge \gstar^2 N
%\]
%\[
%{\betastar^2 } \ge \binom{K_m+1}{2} \frac{q}{N}
%\]
which follows from Lemma \ref{Kbound} since $\gamma_m \le \betastar.$
\end{IEEEproof}

\begin{lemma}[Proof of \eqref{akZ}]
\[
\prob(\kappa(|\survive (\node, \launchid(\node)+K_v)|+1)  < Z_m N )\le e^{-\frac{3}{8}\gstar^2 N}
\]
\end{lemma}
\begin{IEEEproof}
Now, by definition of $\gamma_v$ we have  $\expectation |\survive (\node, (\launchid(\node)+K_v\Tnode)|= \gamma_v \bgstar \betastar$
and $\expectation (\kappa( |\survive (\node, (\launchid(\node)+K_v)\Tnode|+1))= Z_a N.$
Lemma \ref{chernoffsum} yields
\[
\prob\big(|\survive (\node, (\launchid(\node)+K_v)\Delta_t)| - \expectation  |\survive (\node,( \launchid(\node)+K_v)\Delta_t)| \le  -\kappa^{-1} (Z_a - Z_m)\big)
\le e^{-\frac{3}{8}\frac{(Z_a-Z_m)^2 N^2}{\kappa^2\binom{K_v+1}{2}q}}
\]
Thus, the desired result will follow from
\begin{align*}
\frac{ (Z_a - Z_m)^2 N^2 }{\kappa^2 \binom{K_v+1}{2} q} \ge \gstar^2 N
\end{align*}

We have
\begin{align*}
\frac{1}{\kappa} (Z_a - Z_m)
&=
- \bgstar \ln \Bigl(1 - \frac{(\gamma_m-\gamma_a)  \xi_N  }{1 - \xi_N  \gamma_a }\Bigr)
\\
& \ge
\bgstar  \frac{(\gamma_m-\gamma_a) \xi_N}{1-\gamma_a \xi_N}
\\
&= \frac{\gstar \bgstar^2 \betastar}{ 1- \xi_N \bgstar^2\betastar}
\\
&\ge \gstar \bgstar^2 \betastar
\\
&> \gstar \gamma_v
\end{align*}
and Lemma \ref{Kbound} now gives the desired result.
\end{IEEEproof}

Although not needed in the proof, a similar argument shows that the probability of $z_s$ reaching $z_v$ is exponentially small in $N$.

\section{Chernoff Bounds}

In this section we prove some standard inequalities in a form convenient for the proofs.

%\begin{lemma}\label{chernoff}
%Let $b$ be a Bernoulli random variable with $\prob ( b=1 ) = p$ and let $\eta$ be an exponential random variable of rate 1.
%Assuming $s \in (0,3/4)$ we have
%\begin{align*} 
%\expectation e^{s (b - p)}  &\le e^{(1-p) \frac{s^2}{2}} \\
%\expectation e^{-s (b - p)}  &\le e^{(1-p) \frac{2 s^2}{3}} \\
%\expectation e^{s (\eta-1)}  &\le e^{\frac{s^2}{1}} \\
%\expectation e^{-s (\eta-1)}  &\le e^{\frac{s^2}{2}} 
%\end{align*}
%\end{lemma}
%\begin{IEEEproof}
%For any real $s$ we have \( \expectation e^{s (1-b)} =  (1 + (1-p) (e^{s}-1)) \).
%For $s\in(-1/2,1/2)$ we have $(1 + (1-p) (e^{s}-1)) \le e^{(1-p)(e^s-1)}$ which then yields
% \( \expectation e^{-s (b-p)}  \le e^{(1-p)(e^s-1-s)}.$
% The first two inequalities now follow by bounding $e^s-1-s$ from below.
% Combining the two inequalities and noting that the argument can be applied to $1-b$ we obtain
% \[
% \expectation e^{s (b - p)}  \le e^{\min\{p,1-p\} \frac{2 s^2}{3}}
% \]

\begin{lemma}\label{chernofforig}
Let $b$ be a Bernoulli random variable with $\prob ( b=1 ) = p.$
Assuming $s \in (0,3/4)$ we have
\begin{align*} 
\expectation e^{s (b - p)}  &\le e^{(1-p) \frac{s^2}{2}} \\
\expectation e^{-s (b - p)}  &\le e^{(1-p) \frac{2 s^2}{3}} \\
\end{align*}
\end{lemma}
\begin{IEEEproof}
For any real $s$ we have \( \expectation e^{s (1-b)} =  (1 + (1-p) (e^{s}-1)) \).
For $s\in(-3/4,3/4)$ we have $(1 + (1-p) (e^{s}-1)) \le e^{(1-p)(e^s-1)}$ which then yields
 \( \expectation e^{-s (b-p)}  \le e^{(1-p)(e^s-1-s)}.\)
 The two inequalities now follow by bounding $e^s-1-s$ from below.
 \end{IEEEproof}
 
 Combining the two inequalities and applying them to $1-b$ we have the following corollary.
 \begin{corollary}\label{chernoff}
 If $s \in [-3/4,3/4]$ then
 \begin{align*}
  \expectation e^{s (b - p)}  &\le e^{\min\{p,1-p\} \frac{2 s^2}{3}}
    \end{align*}
 \end{corollary}

\begin{lemma}\label{chernoffsum}
%Let $b_k, k=0,1,...,K$ be Bernoulli random variables where $b_k$ is independent of $b_0,...,b_{k-1}$ given $p_k$ (which may depend on $b_0,...,b_k$)
%with $\expectation b_k = p_k,$ let $B = \sum_{k=0}^K {b}_k.$ 
Let $b_k, k=0,1,...,K$ be independent Bernoulli random variables with $\expectation (b_k)=p_k.$ 
%is independent of $b_0,...,b_{k-1}$ given $p_k$ (which may depend on Let $B = \sum_{k=0}^K {b}_k.$  Then
Then
\begin{align}
\prob (B - \expectation B  \ge \eta)  & \le  e^{ -\frac{3\eta^2 }{8 M}} \label{Bernup}
\\
\prob (B - \expectation B  \le \eta)  & \le  e^{ -\frac{3\eta^2 }{8 M}} \label{Berndown}
\end{align}
for any $M \ge \sum_{k=1}^K \min\{p_k,1-p_k\}$ and $\eta \le M.$
\end{lemma}
\begin{IEEEproof}
Using Corollary \ref{chernoff} we have for any $s \in [-3/4,3/4],$
\[
\expectation e^{s (B - \expectation B)}  \le e^{ M \frac{2 s^2}{3}}\,.
\]
Since $\eta \le M$ we now have from the Markov inequality
\begin{align}
\prob (B - \expectation B  \ge \eta)  \le  e^{-s\eta} e^{ M \frac{2 s^2}{3} } \le  e^{ -\frac{3\eta^2 }{8 M}} \label{Bernup1}
\end{align}
where the last step follows by choosing $s = \frac{3 \eta}{4 M}.$ Similarly, we obtain
\begin{align}
\prob (B - \expectation B  \le -\eta)  \le  e^{s\eta} e^{ M \frac{2 s^2}{3} } \le  e^{ -\frac{3\eta^2 }{8 M}} \label{Berndown1}
\end{align}
where the last step follows by choosing $s = -\frac{3 \eta}{4 M}.$
\end{IEEEproof}

\section{\label{appMD1}}

Let $Q_1,Q_2,..$ be i.i.d.  exponential random variables with rate $\gamma < 1.$ 
Let $f(x)$ denote the probability that $x + \sum_{i=1}^m (Q_i  - 1) > 0$ for all $m =0,1,2,...$
Let us extend the definition of $f(x)$ by setting $f(x) = 1$ for $x\in [-1,0).$ 
It follows that for $x > 0$ the function $f(x)$ is the unique fixed point of the map
\(   g \rightarrow {\cal I}(g) \)
defined for bounded non=negative non-increasing functions on $[-1,\infty)$  by
\[
{\cal I}(g)(x) = \begin{cases}
g(x) & x \in [-1,0) \\
 \int_{x-1}^\infty \gamma e^{-\gamma (u - (x-1))} g(u) du & x \ge 0
\end{cases}
\]
It is an easy exercise to show that iterating $\cal I$ on $g$  converges to a unique solution $\gstar$
that depends only on $g(x), x\in [-1,0).$  Furthermore, the solution is monotonic in $g(x), x \in [-1,0),$ i.e.,
given $g_1(x) \le g_2(x), x\in [-1,0),$ it follows that $\gstar_1(x) \le \gstar_2(x).$
We have $f(x) = \gstar(x)$ for $g(x)=1, x\in [-1,0).$ 

Let $\nu$ be the unique solution to $e^{\nu} = 1 + \frac{\nu}{\gamma}.$  
We claim that if $g(x) = e^{-\nu x}, x \in [-1,\infty),$ then $\gstar = g.$
In other words, $e^{-\nu x}$ is a fixed point of ${\cal I}.$
Indeed, for $x\ge 0,$
\begin{align*}
 \int_{x-1}^\infty \gamma e^{-\gamma (u - (x-1))} e^{-nu u} du  
&=
 \int_{0}^\infty \gamma e^{-\gamma u} e^{-\nu (u+(x-1))} du  \\
&=
\frac{\gamma}{\gamma+\nu} e^{\nu} e^{-\nu x} \\
&=
e^{-\nu x}
\end{align*}
By the monotonicity of $\gstar$ as a function of $g(x), x\in [-1,0),$ we now have the following result 
\[
e^{-\nu} e^{-\nu x} < f(x) < e^{-\nu x}\,.
\]

\bibliographystyle{ACM.bst}
\bibliography{AlgorithmWred}

\begin{thebibliography}{1}

\bibitem{Luby19}
{\sc Luby, M.}
\newblock Repair rate lower bounds for distributed storage.
\newblock {\em Accepted to IEEE Transactions on Information Theory\/} (Jan.
  2021).

\bibitem{Luby16}
{\sc Luby, M., Padovani, R., Richardson, T.~J., Minder, L., and Aggarwal, P.}
\newblock Liquid cloud storage.
\newblock {\em ACM Trans. Storage 15}, 1 (Feb. 2019).

\bibitem{saaty1961elements}
{\sc Saaty, T.}
\newblock {\em Elements of Queueing Theory: With Applications}.
\newblock McGraw-Hill, 1961.

\bibitem{MIPSA1985}
{\sc Stecke, K.~E., and Aronson, J.~E.}
\newblock Review of operator/machine interference models.
\newblock {\em International Journal of Production Research 23}, 1 (1985),
  129--151.

\end{thebibliography}


\end{document}